\documentclass[aps,prb,preprint,showpacs]{revtex4-2}

\usepackage{epsfig}

\usepackage{amssymb}

\usepackage{scalerel}
\def\msquare{\mathord{\scalerel*{\Box}{strut}}}

\usepackage{subfigure}
\usepackage{float}
\usepackage{color}      
\usepackage[section]{placeins} 
\usepackage{graphicx}
\usepackage{rotating}
\begin{document}

\title{Boron carbide under torsional deformation: evidence of the formation of chain vacancies in the plastic regime}


\author{Amrita Chakraborti$^{1,2}$}
\email{amrita.chakraborti@polytechnique.edu}
\author{Antoine Jay$^3$}
\author{Olivier Hardouin Duparc$^1$}
\author{Jelena Sjakste$^1$}
\author{Keevin Béneut$^2$}
\author{Nathalie Vast$^1$}
\email{nathalie.vast@polytechnique.edu}
\author{Yann Le Godec$^2$}
\affiliation{$^{1}$ Laboratoire des Solides Irradi\'es, CEA/DRF/IRAMIS, \'Ecole Polytechnique, CNRS, Institut Polytechnique de Paris, 91120 Palaiseau, France}
\affiliation{$^{2}$ Institut de Minéralogie, de Physique des Matériaux et de Cosmochimie (IMPMC), Sorbonne Université, UMR CNRS 7590, Muséum National d'Histoire Naturelle, IRD UMR 206, 4 Place Jussieu, 75005 Paris, France}
\affiliation{$^3$ Laboratoire d'analyse et d'architecture des syst\`emes, CNRS, 31031 Toulouse c\'edex 4, France}

\begin{abstract}
We report a combined experimental and theoretical study of boron carbide under stress/deformation. 
A special rotating anvil press, the rotating tomography Paris Edinburgh cell (RotoPEC), has been used to apply torsional deformation to boron carbide under a pressure of 5~GPa at ambient temperature. 
Subsequent damages and point defects have been analysed at ambient pressure by energy dispersive X-ray microdiffraction at the synchrotron and by Raman spectroscopy, combined with calculations 
based on the density functional theory (DFT). 
We show that apart from the signals due to B$_4$C, new peaks appear in both characterisation methods. The DFT calculations of atomic structures and phonon frequencies 
enable us to attribute most of the new peaks to boron vacancies in the intericosahedral chains of boron carbide. Some of the Raman spectra also show three peaks 
that have been attributed to amorphous boron carbide in the literature. 
Deformed boron carbide thus shows small inclusions of clusters of boron carbide with chain vacancies, and/or small zones interpreted as amorphous zones. 
\end{abstract}

\maketitle

\section{Introduction}
\label{sec:intro}

The properties of boron carbides under stress/deformation have been a long-standing puzzle. On one hand, B$_4$C has outstanding static mechanical properties with, notably, the Vickers hardness ranging from 38 GPa up to 45 GPa in single crystals~\cite{Herrmann:2013}. But on the other hand, its behaviour under dynamical loading is very different. Though it has the highest Hugoniot elastic limit (HEL) among ceramic materials, around 15-17~GPa~\cite{Thevenot:1990, Fanchini:2006, Johnson:1999}, the shear strength in the shocked state rapidly decreases beyond the HEL, resulting in premature failure of the material as the shock stress reaches a threshold value of ~20 GPa~\cite{Domnich:2011}. 

\begin{figure}[t]
\begin{center}
\subfigure[][(B$_{11}$C)C-B-C] 
{\includegraphics[height=0.3\textheight,width=0.3\textwidth]{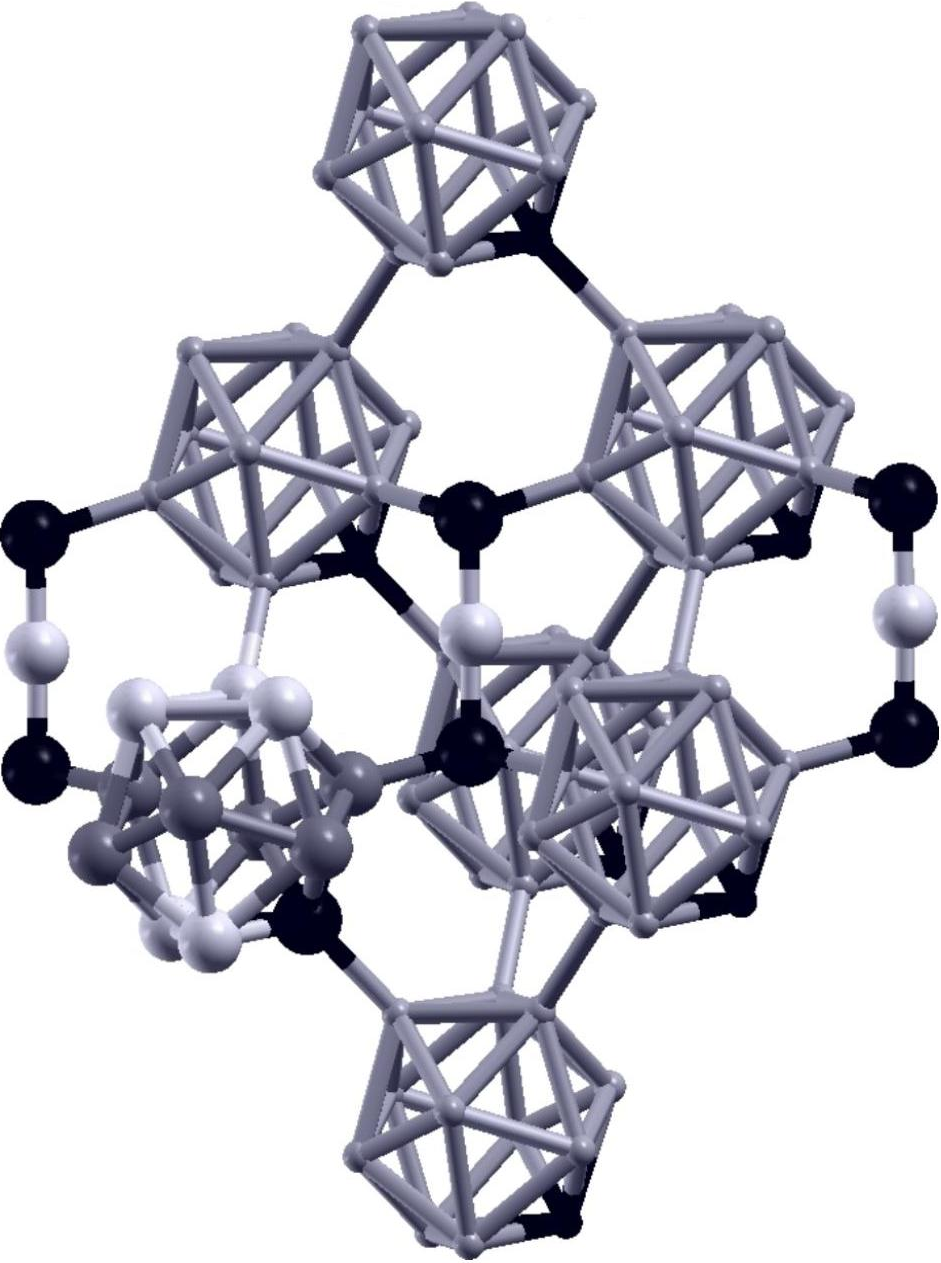} \label{fig:B11C-CBC}}
\subfigure[][C{$\msquare$}C in a B$_4$C matrix] 
{\includegraphics[height=0.3\textheight,width=0.3\textwidth]{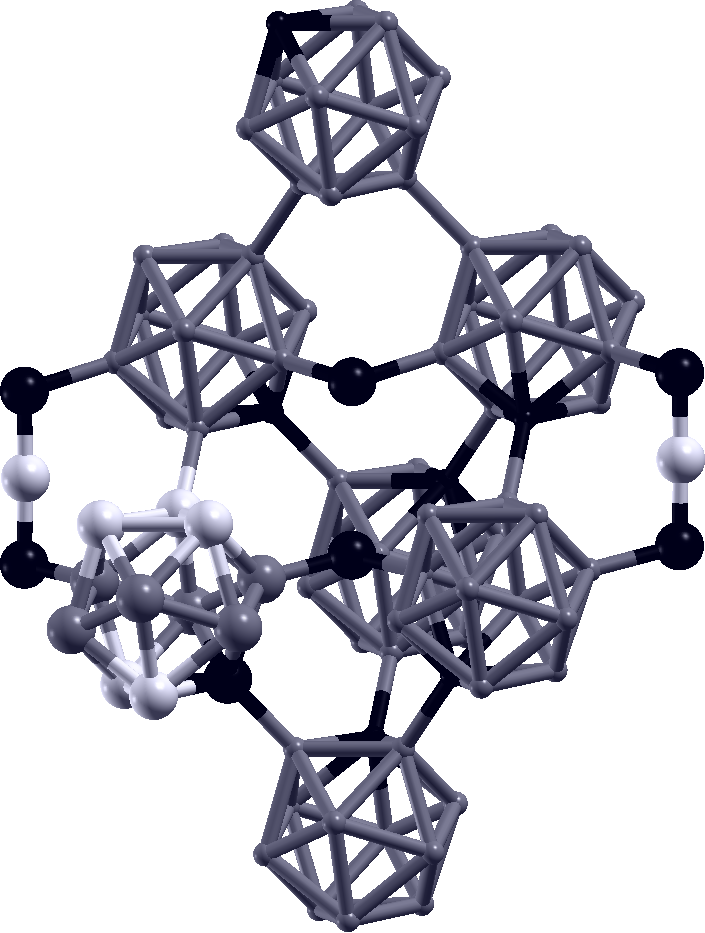} \label{fig:B11C-CBC_defaut_B11C-CVC}}
\subfigure[][(B$_{11}$C)C-C] 
{\includegraphics[height=0.3\textheight,width=0.3\textwidth]{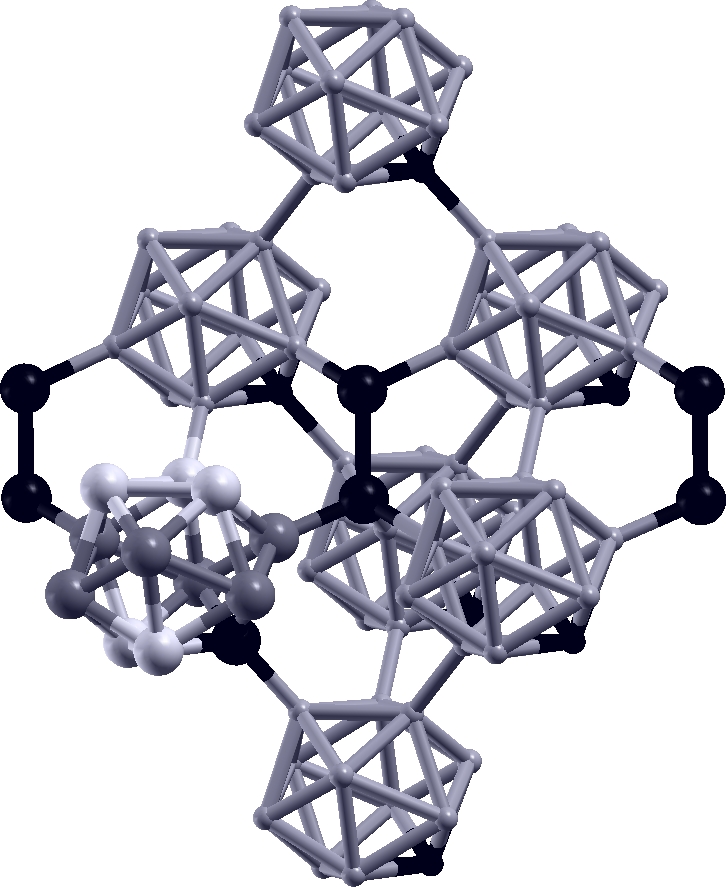} \label{fig:B11C-CC}}
\end{center}
\caption{\label{fig:atomic_structure} Panel a: Atomic structure of B$_4$C boron carbide with ordered (B$_{11}$C) icosahedra. Panel b: Boron vacancy isolated in one of the C-B-C chains of (B$_{11}$C)C-B-C with substitutional polar disorder. Panel c: Ordered (B$_{11}$C)C-C (Ref.~\protect\cite{Jay:2014}). The black balls represent the carbon atoms, the grey balls represent the boron atoms at equatorial sites, while the white balls represent the boron atoms at polar sites and chain centers.}
\end{figure}

Several hypotheses have been put forward to explain the failure of boron carbide under dynamical loading. First, the occurrence of a phase transition has been suggested~\cite{Chakraborti:Note:2021:B4C_Failure}. However, the material after the hypothetical phase transition is in all aspects similar to undeformed boron carbide (figure~\ref{fig:B11C-CBC})~\cite{Vogler:2004}. 
Therefore there is no characterisation of the hypothetical high-stress phase is inadequate. Moreover, in materials that show a polymorphic phase transition under shock, like sapphire~\cite{Reinhart:2006,Zhang:2013,Cao:2017} or GaAs~\cite{Goto:1976,Ono:2018b}, the crystal-to-crystal phase transition also occurs upon reversible loading under static conditions, for instance in a diamond anvil cell~\cite{Sikka:1992}. 
However, X-ray diffraction results have not provided any evidence of a phase transition in boron carbide under reversible loading: the cell parameters decrease monotonically~\cite{Dera:2014}. The behaviour of the Raman peaks and the optical absorption under high pressure have been interpreted as the occurrence of a smooth structural modification in the motif of the boron carbide crystal structure~\cite{Hushur:2016}. Hydrostatic pressure would thus modify the C-B-C chains while leaving the (B$_{11}$C) icosahedral structure intact. 

Finally, another kind of phase transformation has been suggested: the occurrence of shear bands containing amorphous solid have been observed under nanoindentations~\cite{Reddy:2013,Subhash:2013}, 
shock-wave loading experiments~\cite{Chen:2003} 
as well as during one experiment in a diamond anvil cell, upon decompression after loading above 25 GPa~\cite{Yan:2009}. The measured mechanical properties would then be strongly influenced by the amorphous zone~\cite{Ghosh:2012,Subhash:2013}. Nonetheless, the nature of the amorphous solid varies, it could be amorphous carbon or amorphous boron carbide. Its proportion is tiny with respect to the proportion of remaining boron carbide. Moreover, no clear consensus has been reached so far about 
the mechanisms leading to amorphisation in boron carbide nor about the interpretation of experimental and computational results~\cite{Awasthi:2020}. 

Alternatively, some of us have suggested a different mechanism, \textit{via} the formation of point defects, for the loss of mechanical strength beyond the HEL, based on density functional theory (DFT) calculations.
It would result from the formation of vacancies under dynamic loading~\cite{Raucoules:2011}. By computing the formation energy of various neutral vacancies in boron carbide, we have shown that
boron atoms at the centre of the C-B-C chains are the most prone to form vacancies. The drastic consequence of the presence of such chain vacancies in boron carbide driven to the plastic regime
is the formation of new carbon-carbon bonds in the chains: upon application of pressure, new C-C bonds form in the chains, leading to volume variation and ultimately, to the degradation of the material.

The intrinsic concentration of vacancies in boron carbide is too small by itself to lead to mechanical failure~\cite{Yan:2009,Jay:2014}. 
This is compatible with the fact that undeformed boron carbide 
maintains large mechanical strength even under high temperature conditions \cite{Thevenot:1990}. 
However, the amount of point defects in a crystal can increase drastically under plastic deformation, due to the formation and motion of jogs for instance~\cite{Rabier:2010,Hull:2011,Iyer:2014}. 
Therefore, 
our hypothesis is that above the Hugoniot limit, a significant concentration of vacancies appears in the ceramic and a large proportion of them undergoes the 
C$\msquare$C to C-C transformation, where the symbol $\msquare$ stands for the chain boron vacancy. 

However, so far, no experimental proof of the formation of chain vacancies in deformed boron carbide has ever been reported. 
The purpose of the present work is thus to drive boron carbide to the plastic regime in a controlled way and to analyse the subsequent defects and damages thus formed in the material. 
To this end, we report a combined experimental and theoretical study. In the experiments, a special type of rotating anvil press, the rotational tomography Paris-Edinburgh cell (RoToPEC) designed by some of us, has been used to obtain non-hydrostatic torsional stress in boron carbide. The RoToPEC allows us to exercise greater control on the experimental parameters such as temperature and pressure conditions than conventional velocity impact experiments~\cite{Vogler:2004, Holmquist:2006}. 
In the theory, DFT calculations have been used to interpret the subsequent atomic structure, 
 to search for evidence of the chain boron vacancy formation, and to check the conditions under which such C$\msquare$C intericosahedral configurations might be present. 

In section~\ref{sec:expt}, we discuss the materials and methods used in the course of the current work. Next, in section~\ref{sec:results} we report the experimental observations and 
in section~\ref{sec:theory} theoretical results are shown. Experimental and theoretical are compared and discussed in section~\ref{sec:disc}. 
Finally, the conclusions of the work are put forward in section~\ref{sec:conc}.

\begin{figure}[t]
\begin{center}
\includegraphics[width=0.8\textwidth]{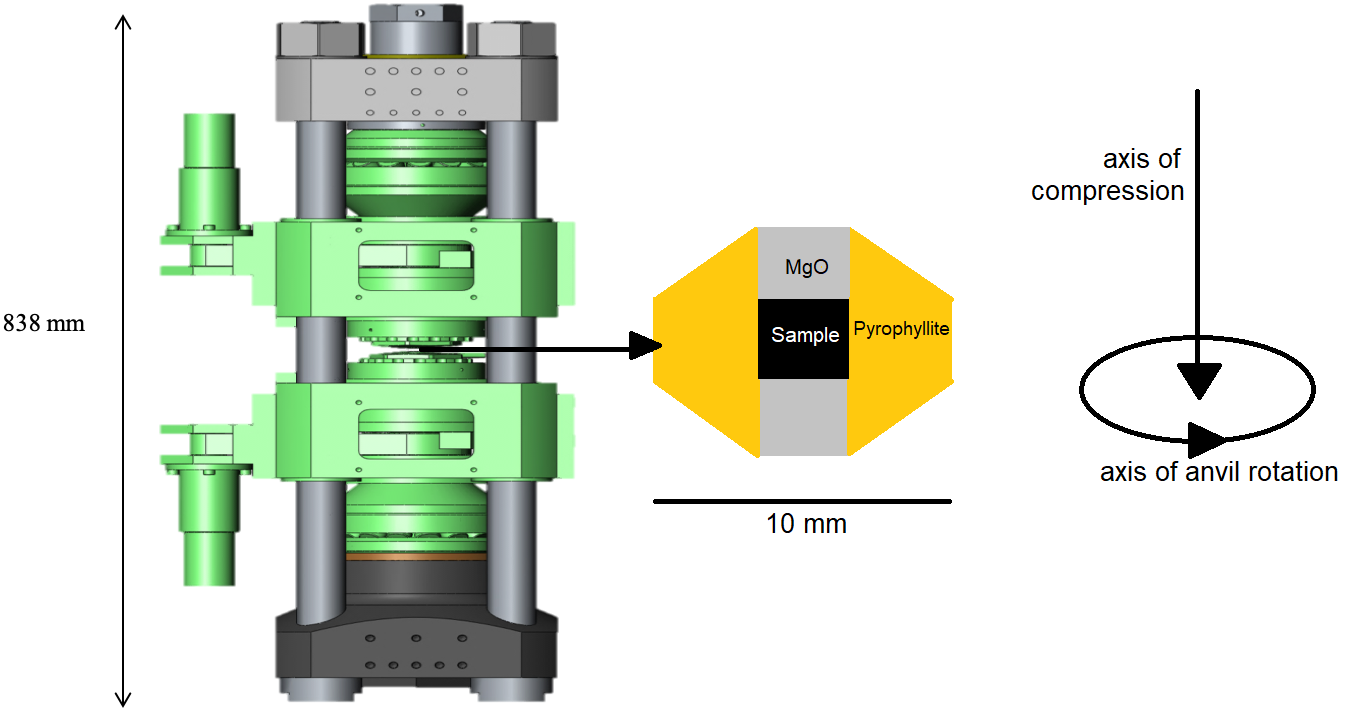}
\end{center}
\caption{\label{fig:gasketrotopec} Schema of the rotating tomography Paris Edinburgh cell (left panel). The left arrow starts in the space between the two anvils, where the fired pyrophillite gasket is inserted (center panel). The boron carbide sample is loaded in the cylindric space inside the gasket, and confined by one MgO plug on each end (see centre panel). In principle, both anvils can rotate to apply torsional deformation to the sample. The direction of the anvil rotation and the direction of compression are shown in the right panel. See details in Ref.~\cite{Philippe:2016}. }
\end{figure}

\section{Materials and methods}
\label{sec:expt}
\subsection{Experimental method} 
\label{subsec:method_expt}

Torsional stress has been applied to compressed boron carbide with the RoToPEC press (figure~\ref{fig:gasketrotopec}). While the detailed mechanical description of this device can be found elsewhere (Ref.~\cite{Philippe:2016}), we briefly mention here that the two opposed anvils can rotate independently under load with no limitation in the rotation angle, through two sets of gear reducers and thrust bearings located at the end of each anvil. The accurate rotation of the anvils is monitored by stepper motors and encoders, with an angular resolution of 0.02°. 

Three such experiments have been performed under a pressure of about 5~GPa, with various degrees of rotation of the anvils : 90\textdegree, 180\textdegree~ and 270\textdegree~ (table~\ref{tab:expparrotopec}). The pressure was determined by the value of the primary pressure on the press, which had been calibrated using \textit{in situ} experiments at synchrotrons with pressure calibrants, similar to the calibration method for the conventional Paris-Edinburgh press \cite{Cherednichenko:2015, Chakraborti:2020, Chakraborti:2021}. In all of our experiments, only the lower anvil was rotated, while the upper one was kept fixed. The speed of rotation of the lower anvil was maintained at 0.01\textdegree/second.
After each experiment, the pressure was released slowly over thirty minutes.

\subsection{Material}
Samples subjected to torsion consisted of commercial boron carbide powder (Alfa Aesar, particle size $\textless$ 10 $\mu$m, 99+ \% purity), that was put inside a fired pyrophyllite gasket (figure~\ref{fig:gasketrotopec}). 
The powder was confined inside the gasket by using one MgO plug on each end. The gasket was then placed inside the RoToPEC and subjected to a pressure of 5~GPa at room temperature, and a torsion of a given angle at the same time. The torsion angle was precisely controlled. We note that in principle, the centre of the gasket has no torsional stress: the further the sample grain(s) are from the gasket centre, the larger the effect of the torsion.

Special attention has been given to the choice of the pressure load in order to reproduce the stress conditions generated in a shock wave experiment with high velocity impact as closely as possible~\cite{Domnich:2011}. 
The chosen value is based on 
the known value of the Hugoniot elastic limit (HEL) of boron carbide, 15-17~GPa~\cite{Thevenot:1990,Fanchini:2006,Johnson:1999}, 
and on the fact that our calculations with density functional theory in the generalized gradient approximation (DFT-GGA) have shown that when boron carbide is heated to a temperature of 2000~K, 
the computed thermal expansion is equivalent to an internal pressure of -12.2~GPa (see e.g.~Ref.~\cite{Jay:2019}). 
Thus, whenever, in a shock wave experiment, a dynamical stress of 17~GPa (the HEL of boron carbide) is applied and the elevation of temperature is up to 2000~K~\cite{Pavlovsky:1971,Dandekar:2001,DeVries:2020},
the equivalent state of stress at ambient temperature amounts to (17-12.2)~GPa~=~4.8~GPa~\cite{Jay:2019}. Therefore, a value of 5~GPa during the generation of the torsion at ambient temperature is expected to be a good approximation of the non-hydrostatic conditions as those near the HEL for boron carbide~\cite{Chakraborti:Note:2021:defect_annealing}. 

\subsection{Post-mortem sample characterisation}

Samples have not been directly recovered after the pressure release.
Due to the nonuniform radial deformation of the sample, it was decided to keep the sample inside the pyrophillite gasket for the post mortem characterisation. This avoids the mixing of undeformed and deformed powder and 
enables us to characterise the defects as a function of the radial distance.

\begin{table}[t]
\hspace*{-1.1cm}
\begin{tabular}{ccccc}
\hline
Expt. name & Rotation (\textdegree) & EDXRD      & Raman      & Summary \\ \hline 
RP~1       & 270                    & changes    & changes    & C$\msquare$C clusters detected \\  
RP~2       & 180                    & no changes & changes    & no C$\msquare$C configurations detected    \\ 
RP~3       &  90                    & no changes & no changes & no C$\msquare$C configurations detected  \\ 
\hline 
\end{tabular}
\caption{\label{tab:expparrotopec} Rotation angles used for the deformation experiments done using the RoToPEC and summary of the characterisations with respect to reference B$_4$C. 
}
\end{table}

Indeed, in principle, apart from what is induced by the (reversible) pressure load, no deformation is expected at the centre of the assembly, while deformation maxima
is expected to occur
on the sample edge near the gasket. In between the centre and the edge of the sample, the deformation is expected to be a function of the radial distance,
as is the torque produced $\tau$, whose expression is $\tau = r \times F$, $r$ being the radial distance and $F$ is the applied force. 

Two complementary characterisations have been performed: energy dispersive X-ray diffraction and Raman spectroscopy. 

\subsubsection{Energy dispersive X-ray diffraction}

\begin{figure}[t]
\begin{center}
 \subfigure[] {\includegraphics[width=0.7\textwidth]{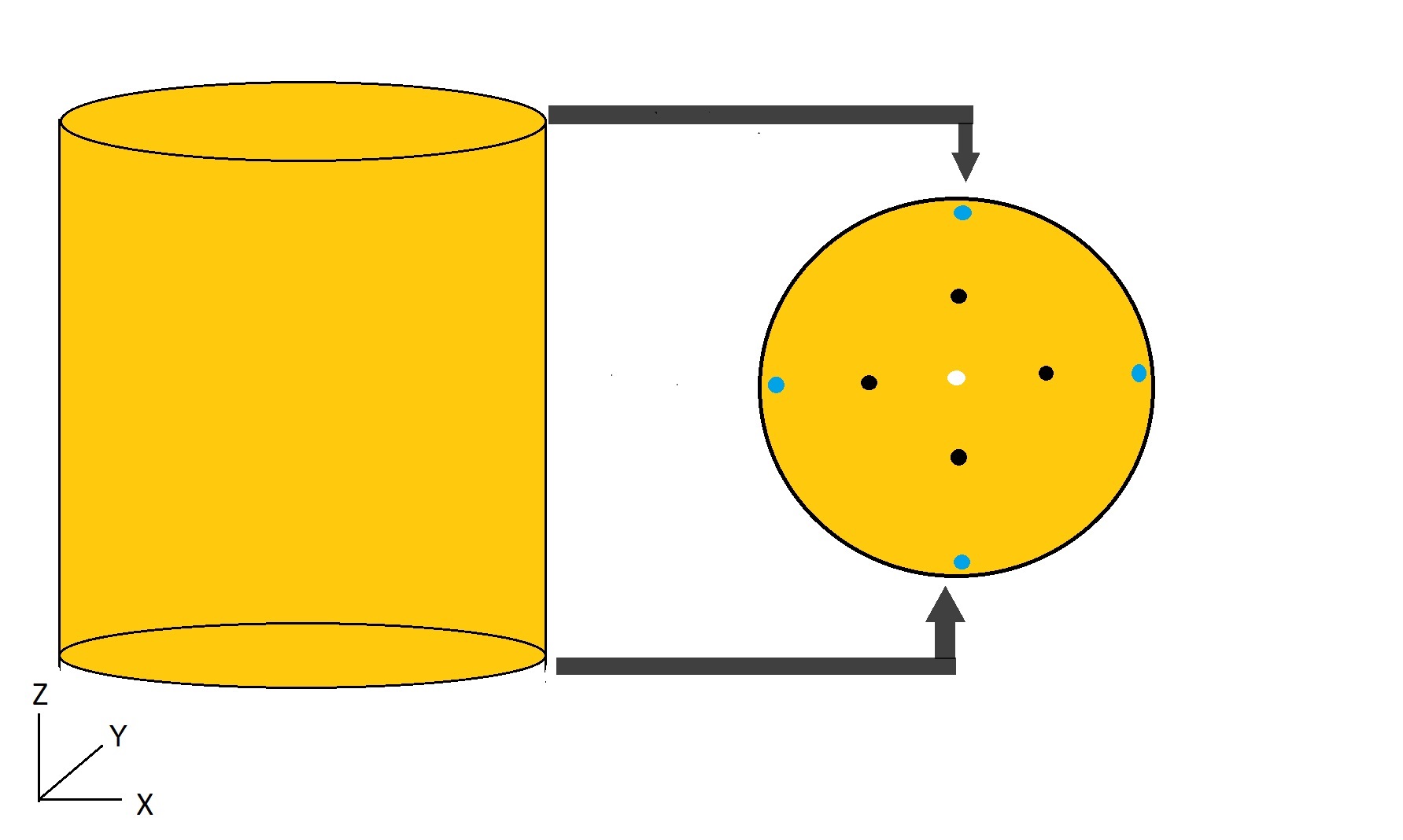} \label{Ramanpositions}}
 \subfigure[] {\includegraphics[width=0.7\textwidth]{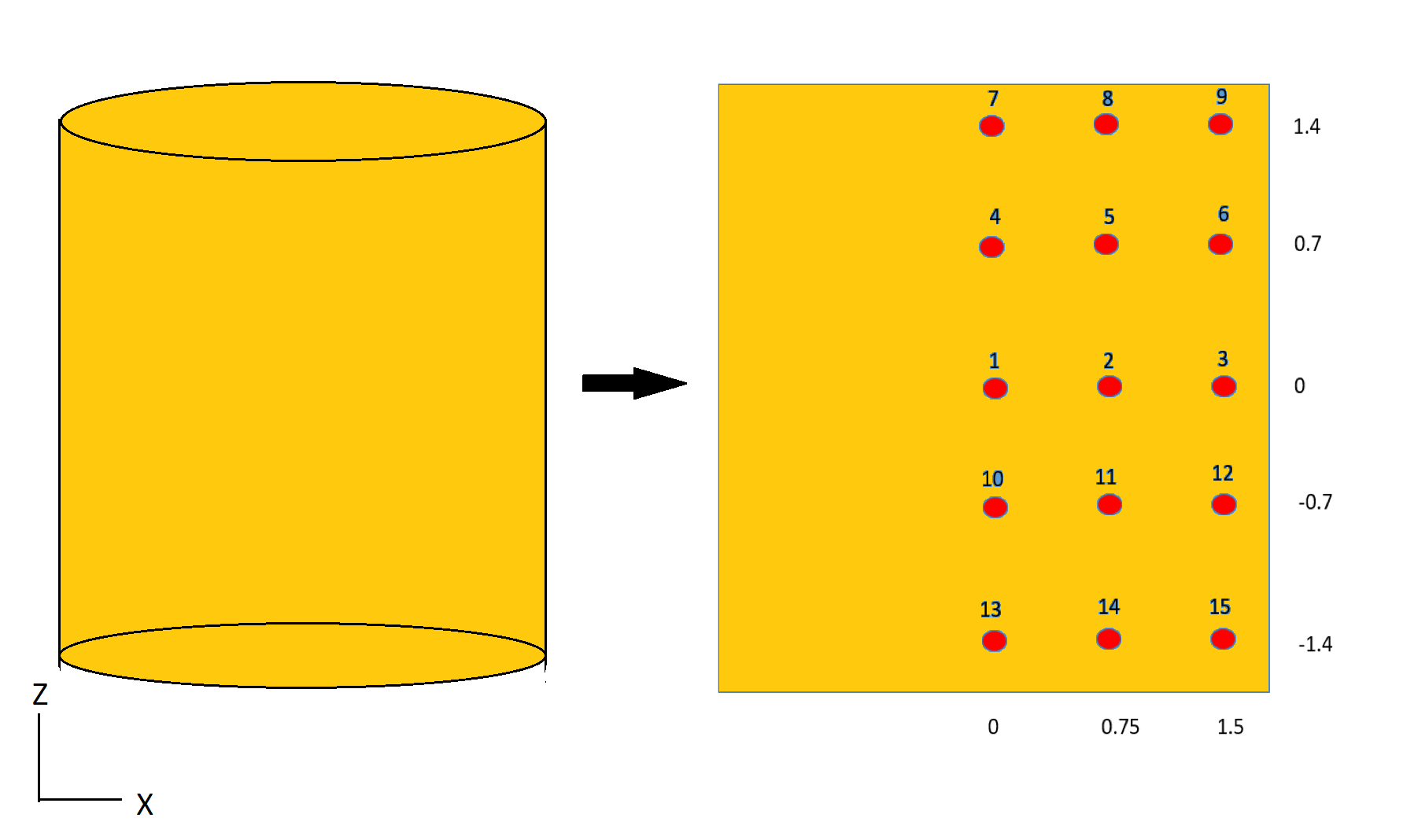} \label{XRDpositions}}
\end{center}
\caption{\label{fig:XRDpositions} (a) For one sample, the nine positions on the top and bottom of the gasket - filled with boron carbide - where Raman spectroscopy has been performed (blue, back, white circles). (b) Positions in the cylindrical space of the gasket where EDXRD has been performed (red colour). The (0,0) coordinates refer to the centre of the sample (equivalent to the gasket centre), marked with the label ``1". The deformation is supposed to be minimum at the centre, while the deformation maxima are expected at $z$= -1.4, given by the labels ``15" since only the lower anvil of the RoToPEC was rotated during the deformation experiment. 
Point coordinates are given in~$mm$, with the sample/gasket centre as the origin. The rectangular shape of the sample results from the projection of the cylindrical space of the gasket filled with boron carbide, onto the $xz$ plane. 
}
\end{figure}

In the first step of the characterisation process, the
deformed samples were observed at the PSICHE beamline in the SOLEIL synchrotron, using energy dispersive X-ray microdiffraction (EDXRD). This
allowed us to characterise different points in the sample volume locally, using the powerful synchrotron radiation without the requirement of any sample recovery from the gasket. Figure~\ref{XRDpositions} shows the locations of all of the points on a cross-sectional view of the sample volume where EDXRD has been performed. 
The positions were chosen assuming that the deformations in the sample would be radially symmetric.

The white beam in SOLEIL has an energy range spanning from 15~to~80~keV and it is focused to a size of 25~$\mu$\textit{m} in the vertical direction and collimated to 50~$\mu$\textit{m} in the horizontal direction of the sample.
This limits the sample size that was scanned at each position, thus ensuring local characterisation when compared to the total sample volume (3.5~\textit{mm} diameter).

The sample cross-section has been mapped using EDXRD at 15 different points of each of the three samples, leading to 45 different EDXRD patterns. Due to space constraint, only the most relevant data 
is shown in the next section.
In the following, the EDXRD data has been converted to the corresponding peaks for Cu~K-$\alpha$ radiation, in order to compare them with the XRD pattern of the initial undeformed boron carbide powder. 

\subsubsection{Raman spectroscopy}

In the second step of the characterisation process, the same gaskets were prepared for the Raman spectroscopy. Each of the gasket was embedded in a cold resin made by mixing two parts by weight of acrylic KM-U and one part by weight of methyl methacrylate. 
The sample never came in direct contact with the resin: the resin only covered the MgO plugs and the pyrophyllite of the gasket.

The gasket embedded in the resin was then polished using a macropolishing device (MECAPOL P230), 
until the sample surface was exposed on both sides. 
The polishing was done as gently as possibly with a 120 grit sandpaper at a speed of 100 revolutions per minute.

The Raman spectroscopy was then performed on both the upper and lower exposed surfaces of the samples. Nine spectra were taken on each side of a sample, leading to 54 spectra in total for both sides of the three samples (figure~\ref{Ramanpositions}). The Raman spectroscopy was done in back-scattering geometry, using a Horiba Jobin Yvon HR800 Raman spectrometer with $\times$ 10 and $\times$ 20 objectives. The 514.5~\textit{nm} line of an Ar$^{+}$ laser, with 2 $\mu$m beamspot, is used to measure the atomic vibration frequencies in the samples.
Similar to the EDXRD results, only a few of the most pertinent spectra have been reported in the following.

It is important to note that the EDXRD and the Raman spectroscopy have not been done 
on the same areas of the sample. 
The Raman spectroscopy has been collected on the top and bottom circular areas of the sample surfaces (figure~\ref{Ramanpositions}), 
as these were the only areas that could be accessed with optical means without destruction of the sample. The areas where Raman spectroscopy was performed, however, are close to the positions 7, 8, 9 and 13, 14, 15 in figure~\ref{XRDpositions}. 

\subsection{Computational methods}
\label{subsec:methods_calc}

Total energy and phonon calculations have been performed within the density functional theory (DFT))~\cite{Hohenberg:1964,Kohn:1965} and density functional perturbation theory (DFPT)~\cite{Baroni:2001} with the pseudopotential and plane-wave method. Both the 
generalised gradient approximation in the PW91 parameterisation (GGA-PW91)~\cite{Perdew:1992} and the local density approximation (LDA)~\cite{Perdew:1981} have been used, the former for the characterisation of the atomic structures, energetics and calculation of the enthalpy, and the latter for the energetics and for the phonon frequencies at the centre of the Brillouin zone (BZ).

Three kinds of models have been used for the defective atomic structure, with the aim of representing either cluster of defects or isolated defects. 

\subsubsection{Ordered models for boron carbide}
Firstly, cluster of defects have been modelled by one single crystalline phase, repeating periodically the 14- or 15-atom unit cell, to obtain the crystal unit cells of
ordered (B$_{11}$C$^p$)C$\msquare$C, (B$_{11}$C$^p$)C-C and of the reference 
(B$_{11}$C$^p$)C-B-C. The subscript~$^p$ indicates that the carbon atom is in one of the six equivalent atomic positions of the polar site of the icosahedron. In such elementary unit cells, the polar carbon atom is located at the same atomic position from one cell to the other (see \textit{e.g.} figures~\ref{fig:B11C-CBC} and~\ref{fig:B11C-CC}), and the crystal symmetry is monoclinic. To obtain the XRD patterns, the base-centered monoclinic unit cell parameters and atomic positions have been averaged in the trigonal symmetry as reported in~\ref{app:structural_model_details}. The averaging procedure is described in~\ref{app:average_trigonal_unit_cell}.

Phonons at zone centre have been computed for these structural models. Main approximations for the computation of vibrational properties
are the Born-Oppenheimer approximation and the harmonic approximation. 

From the group theory, one expects 12 Raman-active frequencies for B$_4$C in the $R\overline{3}m$ space group, of which seven modes have the $E_g$ symmetry and five modes have the $A_{1g}$ symmetry~\cite{Lazzari:1999}.
Removing the chain centre atom only affects the number of infrared active modes so that 12 Raman-active modes are also expected for (B$_{11}$C$^p$)C$\msquare$C and (B$_{11}$C$^p$)C-C. 
The phonon eigenvectors of the monoclinic cells have then been projected onto the phonon eigenvectors of a crystal with the trigonal symmetry to retrieve the vibrational modes that are Raman active in the average trigonal symmetry. 
We choose the theoretical (B$_{12}$)C-C-C crystal as our trigonal reference. To follow each of the Raman-active modes as closely as possible, we have designed the projection paths given below~\cite{Jay:2015}: 
\begin{itemize}
\item (B$_{12}$)C-C-C $\rightarrow$ (B$_{12}$)C$\msquare$C $\rightarrow$ (B$_{11}$C$^p$)C$\msquare$C for the vacancy;
\item (B$_{12}$)C-C-C $\rightarrow$ (B$_{12}$)C$\msquare$C $\rightarrow$  (B$_{12}$)C-C $\rightarrow$  (B$_{11}$C$^p$)C-C for the diatomic chain;
\item (B$_{12}$)C-C-C $\rightarrow$ (B$_{12}$)C-B-C $\rightarrow$  (B$_{11}$C$^p$)C-B-C for the reference boron carbide.
\end{itemize}

\subsubsection{Models for substitutional disorder in boron carbide}

Secondly, to investigate the effect of the lift of symmetry in the ordered models, and to reproduce numerically the more symmetric R$\bar{3}m$ space group observed experimentally in boron carbide, 3x3x3 supercells that contained 27 icosahedra have been constructed and the substitutional disorder of carbon atoms in the six equivalent atomic positions of the polar site of the icosahedra was reproduced (see~\ref{app:modelling_substitutional_disorder}). This enables us to obtain the trigonal symmetry for all of the XRD peaks, in the sense that only a limited number of peaks have a significant intensity, the intensity of the other ones being vanishingly small. The positions of these peaks correspond nicely to the position of the peaks of the trigonal average of the ordered structure. One exception is the (221) peak of the disordered (B$_{11}$C$^p$)C$\msquare$C phases (table~\ref{tab:XRD_theory}, last raw, 4$^{th}$ column), which is consistent with the fact that (B$_{11}$C$^p$)C$\msquare$C has the largest monoclinic distortion (~\ref{app:structural_model_details}, table~\ref{tab:equi_prop_GGA}). The higher the monoclinic deformation and the Miller indices, the larger the supercell size required in order to be close to the average trigonal symmetry. As the (221) peak does not play a significant role in the discussions of section~\ref{sec:disc}, the 3x3x3 supercell is sufficient for our purpose.

$2 \theta$ values of the disordered phases were then compared to symmetrically averaged values of the ordered phases. They turn out to be only slightly different from corresponding peaks of the ordered phase: the mean squared difference (MSD) between the ordered phase and the phase with polar disorder amounts to~0.036\textdegree~ for (B$_{11}$C)C$\msquare$C (table~~\ref{tab:XRD_theory}, comparison of columns~2~and~4), to~0.024\textdegree~ for B$_4$C (columns~8~and~10) and to~0.015\textdegree~ for (B$_{11}$C)C-C (columns~12~and~14). The MSD is the largest for (B$_{11}$C)C$\msquare$C and the smallest for (B$_{11}$C)C-C, B$_4$C MSD being inbetween, which is consistent with the fact that the monoclinic distortion is the largest for (B$_{11}$C)C$\msquare$C and the smallest for (B$_{11}$C)C-C (~\ref{app:structural_model_details}, table~\ref{tab:equi_prop_GGA}). 
In the following, XRD experiments will be compared with either of the two results, and the quantitative comparison (mean-squared differences) will be estimated with values of calculations with polar disorder.

\subsubsection{Modelling isolated defects in boron carbide}

Third, isolated defects have been modelled by one single (B$_{11}$C)C-C (resp.~(B$_{11}$C)C$\msquare$C) defect in a 3x3x3 matrix of B$_4$C. Out of the 27 chains contained in the simulation cell, 26 were C-B-C ones and the remaining icosahedral space contained one C$\msquare$C configuration (resp.~one C-C chain). Substitutional disorder of the carbon atom in the polar site of the icosahedra has been accounted for. 

\subsubsection{Computational details}

The size of the plane wave basis set has been limited with a cutoff energy of 80 Ry. The BZs of the elemental unit cells have been sampled with a 12×12×12 ~\textbf{k}-point mesh centred at $\Gamma$, whereas the BZs of the supercells have been sampled with a 2x2x2 Monkhorst-Pack mesh~\cite{Monkhorst:1976}.
Metallicity has been treated by the Methfessel-Paxton smearing with a width of 10~mRy~\cite{Methfessel:1989}. All of the lattice parameters and atomic positions have been relaxed and the main equilibrium properties are reported in the appendix, in tables~\ref{tab:equi_prop_GGA}~and~\ref{tab:equi_prop_LDA}. 

Finally, our aim was also to estimate the energy barrier between the C$\msquare$C and C-C configurations. Indeed, the distance between the two carbon of the chain 
can be seen as a reaction path to go from one phase to the other, or from one isolated defect to the other. For the ordered phases, this distance has been fixed and the enthalpy 
calculated while all remaining atomic positions and cell parameters were relaxed. This enabled volume relaxation along the path. 
For the isolated defect, we used instead the climbing image nudge elastic band method~\cite{Pratt:1986,Elber:1987} which enabled us to find the minimum energy path at the volume of the B$_4$C matrix.

\section{Experimental results}
\label{sec:results}

\subsection{Energy dispersive X-ray diffractograms}
\label{subsec:results_EDXRD}

\begin{figure}[t]
\begin{center}
\includegraphics[width=0.7\textwidth]{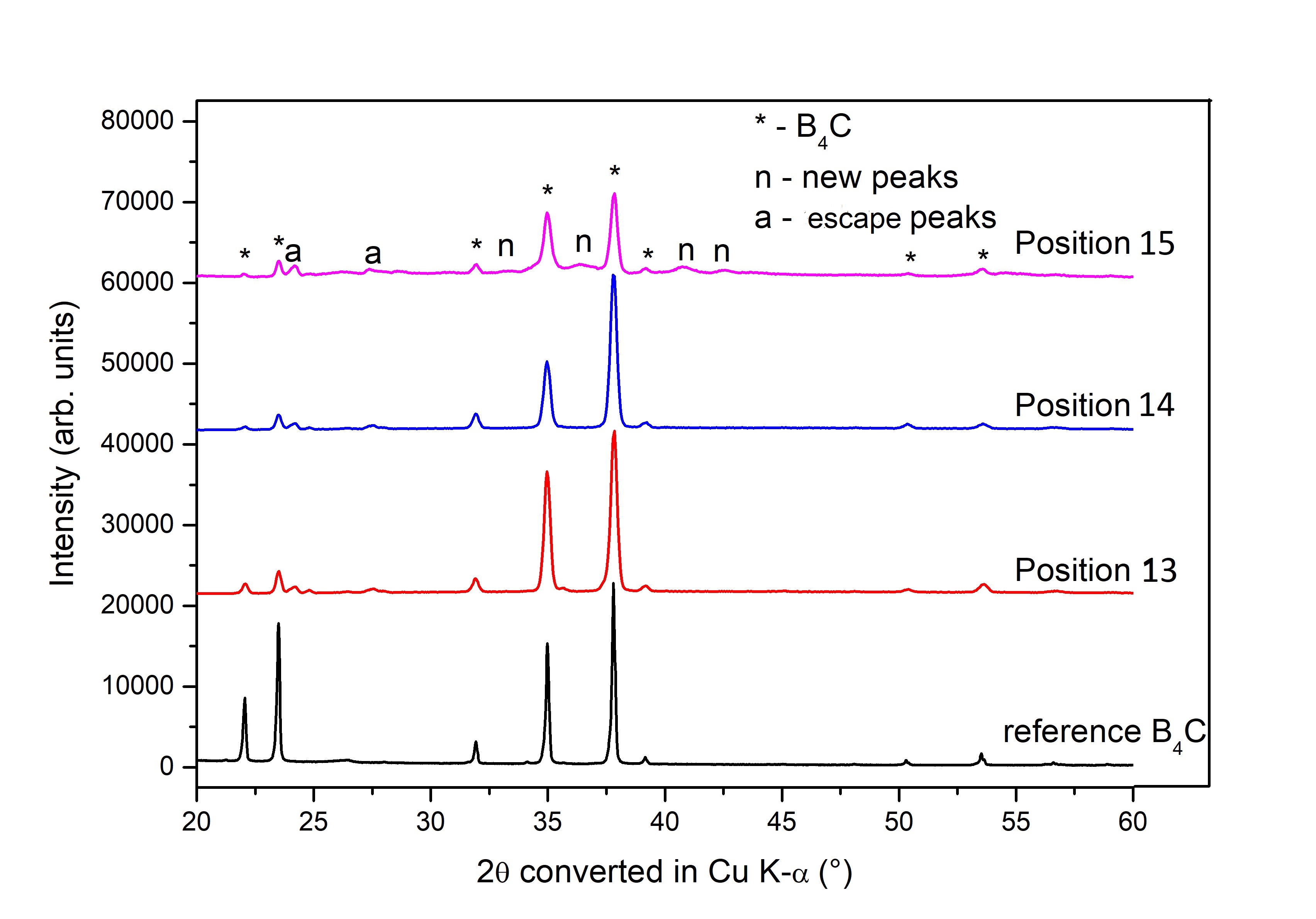}
\end{center}
\caption{\label{fig:RP1} Characterisation at ambient pressure of the sample from the RP~1 experiment. X-ray microdiffraction pattern of the rectangular cross-section of the gasket rotated by~270\textdegree~ under 5~GPa in the RoToPEC (table~\ref{tab:expparrotopec}). Positions~15~and~13 refer to the gasket edge and centre respectively, position~14 being at midway between the centre and the edge (see figure~\ref{XRDpositions}). The black line shows the XRD of the boron carbide powder at ambient pressure before deformation.
}
\end{figure}

With respect to initial undeformed boron carbide (figure~\ref{fig:RP1}, bottom curve), several new diffraction peaks appear in the XRD patterns of the RP~1 experiment when taken on the gasket edge (position~15, upper curve).

Such peaks are absent when taken at the gasket center (position~13) or at midway between center and edge (position~14) (resp. third and second curves from top to bottom).
In the latter cases, the atomic structure of boron carbide is left unchanged.

In the following, the new peaks seen on the sample edge are interpreted as the formation of clusters of defects. 
This set of results is consistent with the expectation that the maximum deformation should occur at the edge of the sample, at the largest distance from the gasket centre. Moreover, the deformation should also be maximum at low $z$ (position 15 in figure~\ref{XRDpositions}) and minimum at high $z$ (position 9 in figure~\ref{XRDpositions}) since only the lower anvil of the RoToPEC was rotated.

\begin{figure}[t]
\begin{center}
 \subfigure[EDXRD on the gasket after the RP~2 expt.] {\includegraphics[width=0.65\textwidth]{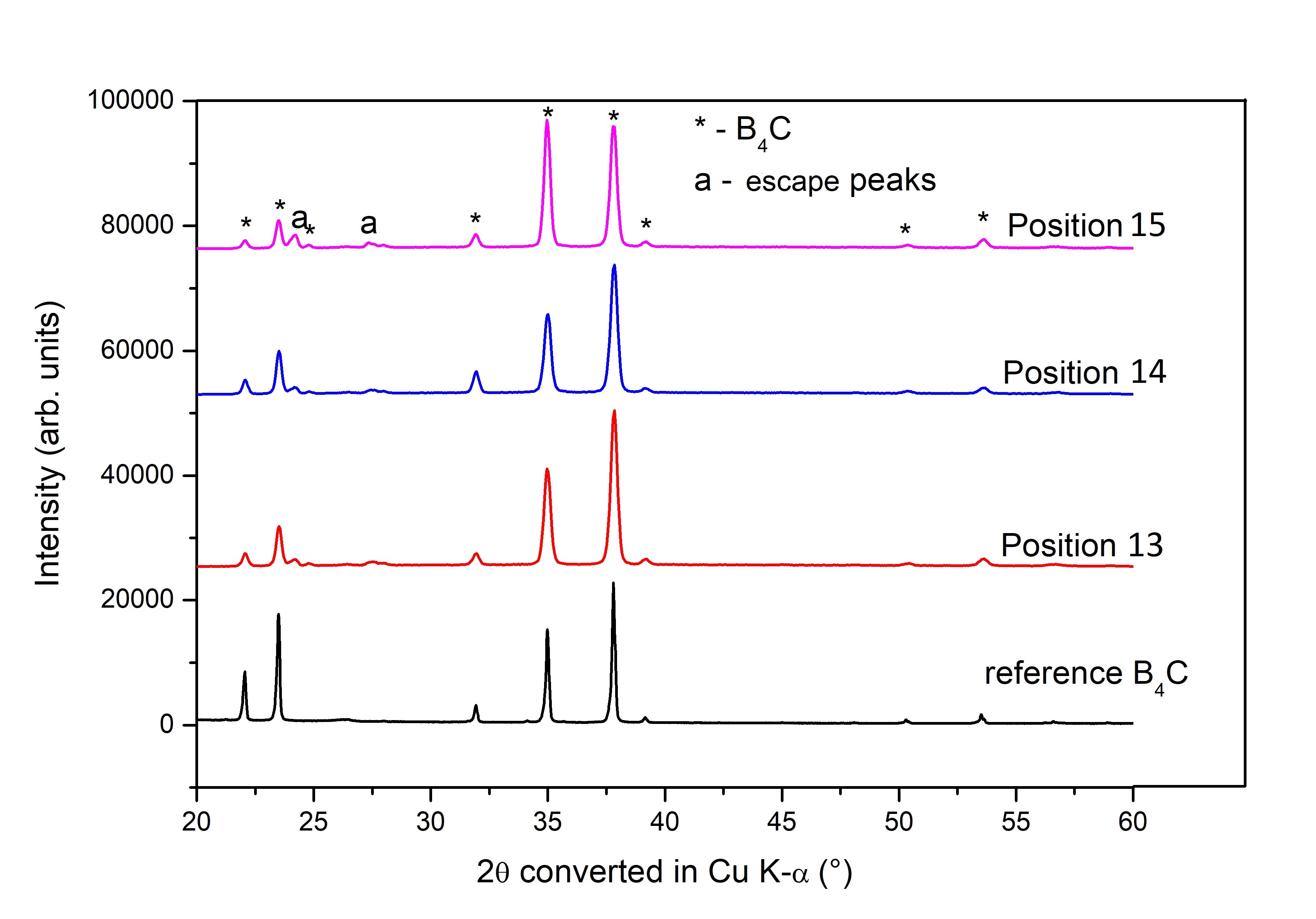} \label{fig:RP2}}
 \subfigure[EDXRD on the gasket after the RP~3 expt.] {\includegraphics[width=0.65\textwidth]{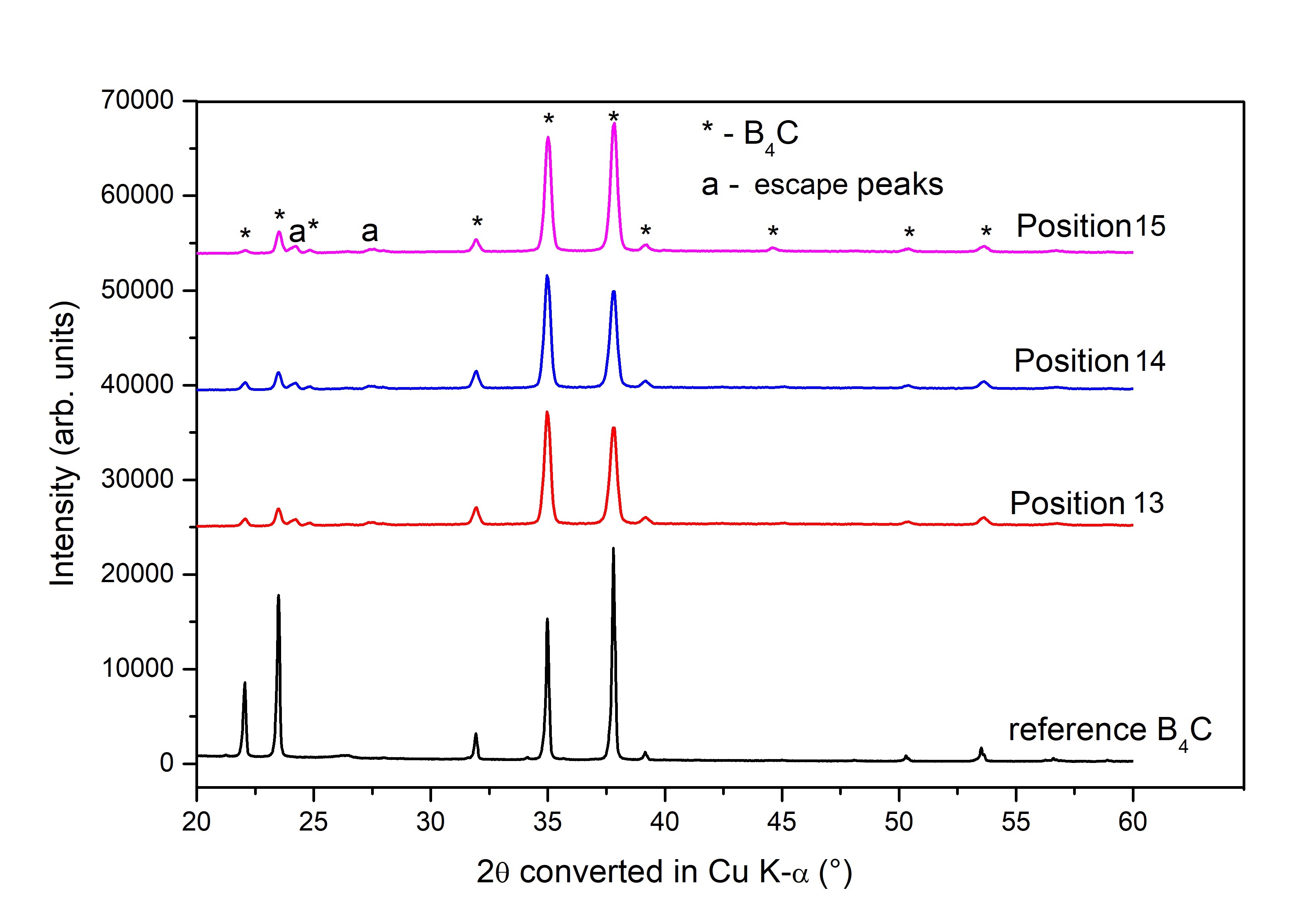} \label{fig:RP3}}
\end{center}
\caption{\label{fig:RP2_and_RP3} X-ray microdiffraction pattern of the rectangular cross-section of the gasket rotated by resp. ~180\textdegree~ (RP~2~experiment, top panel) and~90\textdegree~ (~RP~1~experiment, bottom panel). Same legends as in figure~\protect\ref{fig:RP1}.
}
\end{figure}

In order to demonstrate the effect of the degree of rotation on the samples, the XRD patterns on the same positions as in the RP~1 experiments are shown for the RP~2 and RP~3 experiments defined in table~\ref{tab:expparrotopec} (figure~\ref{fig:RP2_and_RP3}, resp. top and bottom panels). These patterns show no difference with respect to the distance from the centre.
Hence, a minimum deformation threshold is probably needed to activate the transformation mechanism.

Two unknown peaks marked with the label "a" have appeared in all of the spectra at the 2$\theta$ values of 24.2\textdegree~ and 27.4\textdegree.
These peaks appear whatever the torsion angle, even near the centre of the samples where the deformation induced by the torsion is negligible,
as confirmed by the non variation of other peaks in this region. Hence, they are attributed to escape peaks from the germanium detector,
and not to a peak arising because of defects formed in the sample itself. If these peaks are excluded, four peaks are new and all of them coincide with those expected by the theory for
(B$_{11}$C$^p$)C$\msquare$C, as will be shown in Sec.~\ref{sec:disc} below.

\subsection{Raman spectra}

\begin{figure}[t]
\vspace*{-2.2cm}
\begin{center}
 \subfigure[Along direction~1] {\includegraphics[height=0.28\textheight,width=0.5\textwidth]{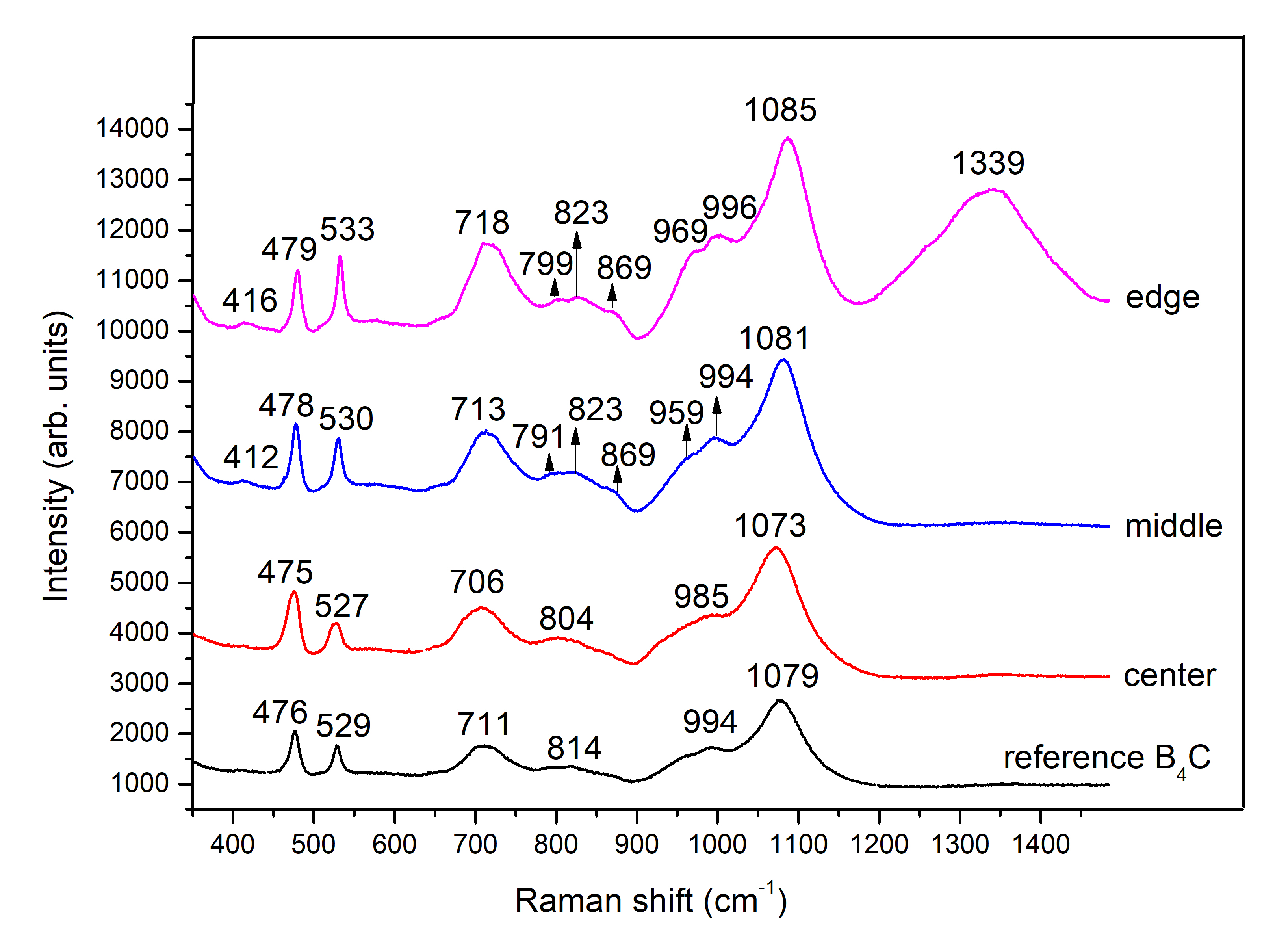} \label{fig:RP1_Raman_2_1}}
 \subfigure[Along direction~2] {\includegraphics[height=0.28\textheight,width=0.5\textwidth]{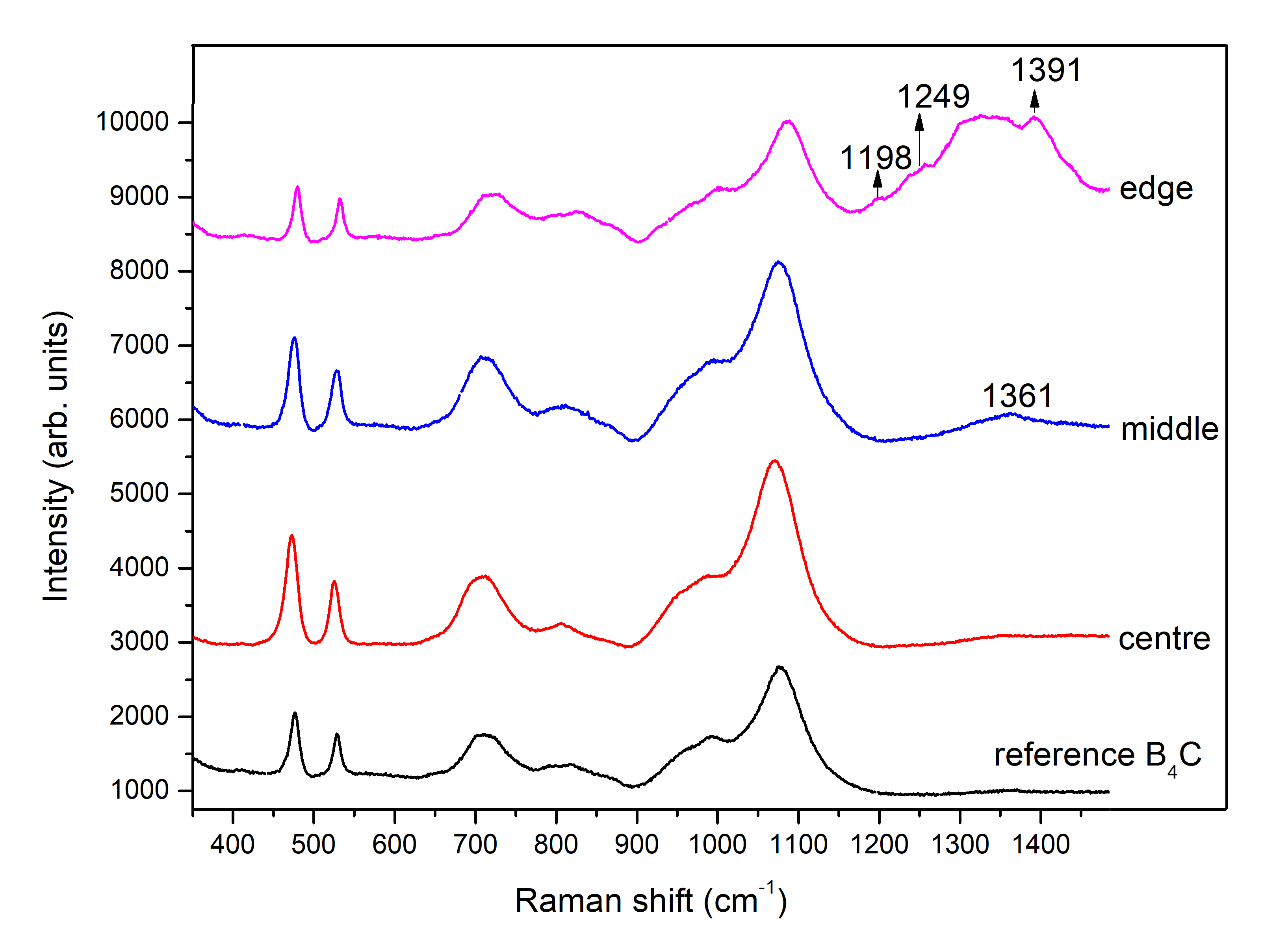} \label{fig:RP1_Raman_1_1}}
 \subfigure[Along direction~3] {\includegraphics[height=0.28\textheight,width=0.5\textwidth]{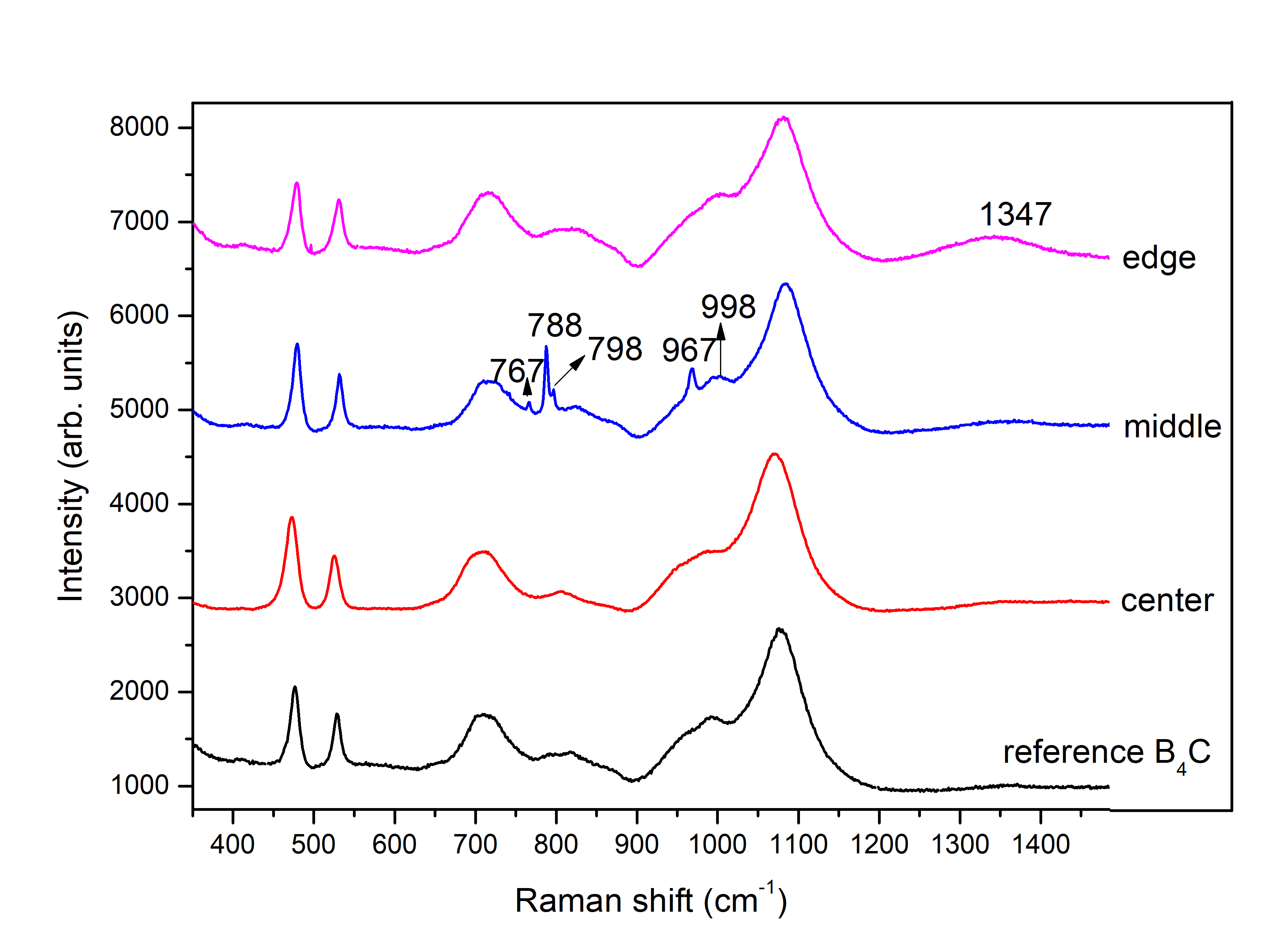} \label{fig:RP1_Raman_1_2}}
\end{center}
\caption{\label{fig:RP1_Raman} \small{The Raman spectra along three distinct directions of the circular sample surface rotated by~270\textdegree~ (RP~1 experiment). The position where the spectrum was taken has been varied from the disc-centre to the disc-edge along the radius. The black line shows the spectrum of initial undeformed boron carbide powder at ambient pressure.} 
}
\end{figure}

For the sample of the RP~1 experiment, the Raman spectra along three distinct directions of the circular surfaces also showed changes when the probing laser beam was moved from the gasket centre towards the gasket edge (figure~\ref{fig:RP1_Raman}).

However, the correlation between spectral changes and the radial distance at which the spectrum has been taken is not as strict as in the case of the XRD patterns. This can be seen in particular
in figure~\ref{fig:RP1_Raman_1_2}, where at midway between the gasket centre and edge, the number of peaks is larger than on the edge, and the additional peaks appear to be more harmonic than the initial undeformed boron carbide peaks.

\begin{figure}[t]
\begin{center}
 \subfigure[The Raman spectrum after the RP~2 experiment] {\includegraphics[width=0.7\textwidth]{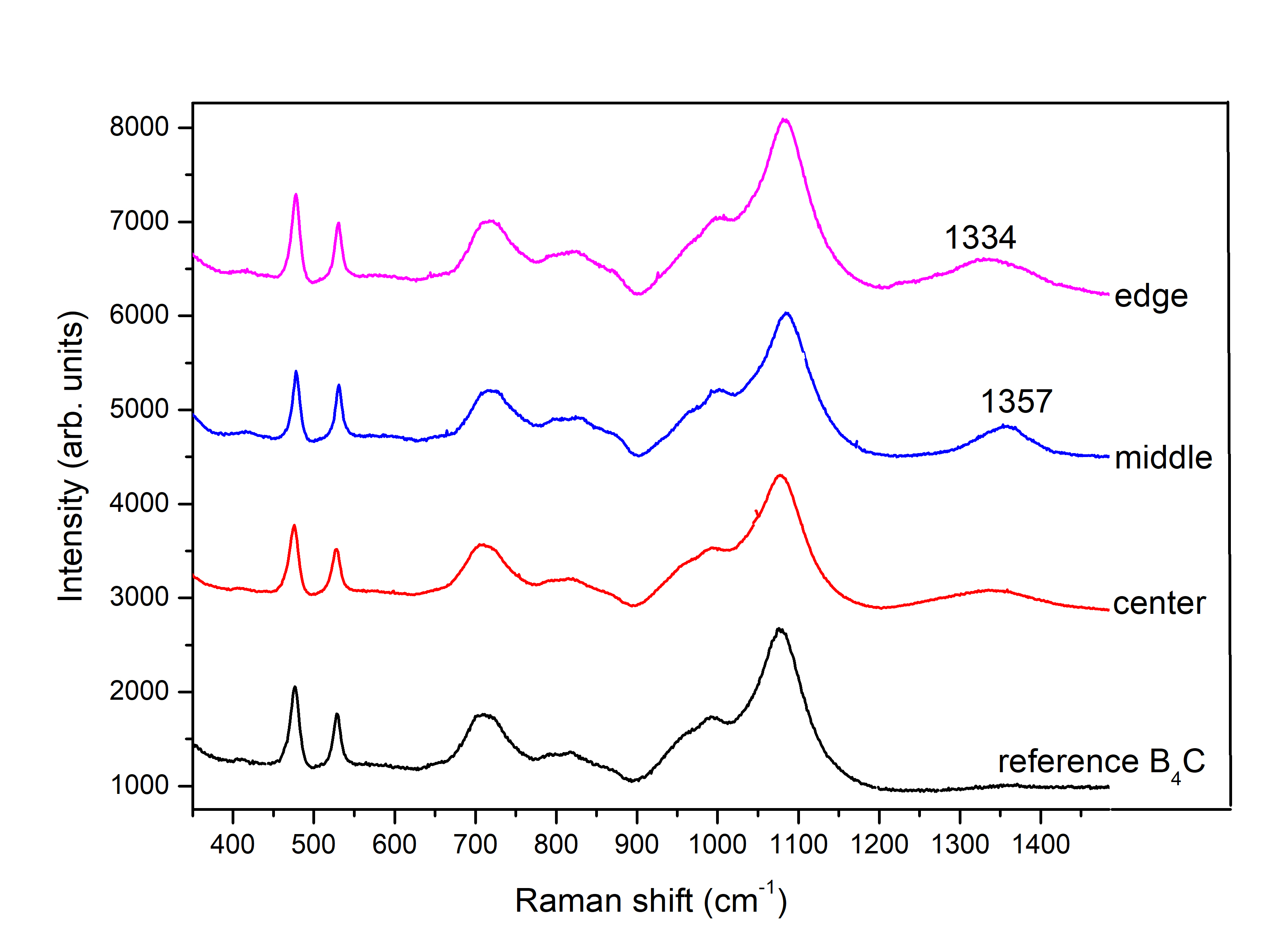} \label{fig:RP2_Raman}}
 \subfigure[The Raman spectrum after the RP~3 experiment] {\includegraphics[width=0.7\textwidth]{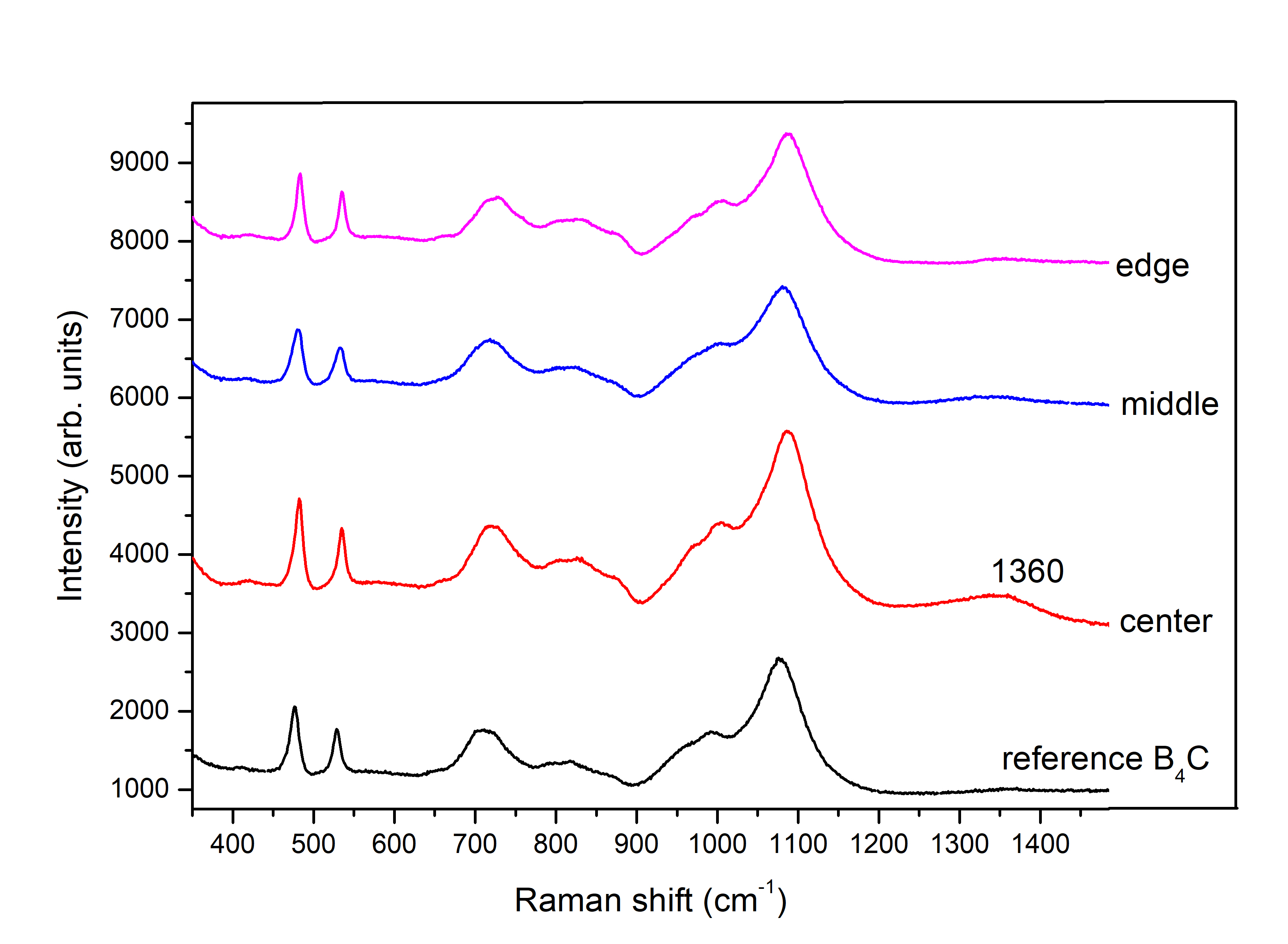} \label{fig:RP3_Raman}}
\end{center}
\caption{\label{fig:RP2_and_RP3_Raman} The Raman spectra of the circular surface of the gasket rotated by resp.~180\textdegree~ (RP~2~experiment) and~90\textdegree~ (RP~3~experiment) in the RoToPEC along one direction. Same caption as in figure~\protect\ref{fig:RP1_Raman}. No amorphous boron carbide signature is observed in RP~3, unlike RP~2. 
}
\end{figure}

Raman spectra of the sample obtained from the RP~2 experiment show some changes (figure~\ref{fig:RP2_and_RP3_Raman}, top panel), while no changes have been observed for the RP~3 experiment for which the torsion is the smallest (panel ~\ref{fig:RP3_Raman}).
These Raman observations underscore the effect of the angle of rotation in creating deformation, and thus defects, in the sample.
The larger the torsion angle, the greater the number of changes in the spectra.

\section{Theoretical results}
\label{sec:theory}

In the following, we look for the spectroscopic fingerprints of the C$\msquare$C configuration in large clusters  or as a defect isolated in a matrix of B$_4$C (fig.~\ref{fig:B11C-CBC_defaut_B11C-CVC}).
We also study the configuration in which the C-C bond is formed and the chain is diatomic (fig.~\ref{fig:B11C-CC}), as this configuration with diatomic chains is expected at large pressure~\cite{Raucoules:2011,Betranhandy:2012}.
It needs to be understood whether, in the experiments, the pressure of 5~GPa applied together with the torsional deformation is expected to lead to the C$\msquare$C $\rightarrow$ C-C transformation once chain vacancies have been formed.

\begin{sidewaystable}
\begin{tabular}{lcrcrcrcrcrcrcrcr}
\hline
                             & \multicolumn{6} {c}{\footnotesize{C$\msquare$C}}  & \multicolumn{4} {c}{\footnotesize{B$_4$C} matrix}   &  \multicolumn{6} {c}{\footnotesize{C-C}}        \\
\footnotesize{$R\overline{3}m$}     & \multicolumn{2} {c}{\footnotesize{Ordered}}       & \multicolumn{2} {c}{\footnotesize{Disordered}}     &\multicolumn{2} {c}{\footnotesize{Isolated defect}} &
                               \multicolumn{2} {c}{\footnotesize{Ordered}}       & \multicolumn{2} {c}{\footnotesize{Disordered}}     &
                               \multicolumn{2} {c}{\footnotesize{Ordered}}       & \multicolumn{2} {c}{\footnotesize{Disordered}}     &\multicolumn{2} {c}{\footnotesize{Isolated defect}} \\
\footnotesize{$hkl$}  &\footnotesize{2$\theta$ (\textdegree)} &  \footnotesize{$I$} & \footnotesize{2$\theta$ (\textdegree)} &  \footnotesize{$I$} & \footnotesize{2$\theta$ (\textdegree)} &  \footnotesize{$I$} &
                \footnotesize{2$\theta$ (\textdegree)} &  \footnotesize{$I$} & \footnotesize{2$\theta$ (\textdegree)} &  \footnotesize{$I$} &
                \footnotesize{2$\theta$ (\textdegree)} &  \footnotesize{$I$} & \footnotesize{2$\theta$ (\textdegree)} &  \footnotesize{$I$} & \footnotesize{2$\theta$ (\textdegree)} &  \footnotesize{$I$}  \\
\hline
\footnotesize{$1\,0\,0$}            &
\footnotesize{19.96} & \footnotesize{(46)}          & \footnotesize{19.87}       & \footnotesize{(7)}         & \footnotesize{19.70} & \footnotesize{(13)}        & \footnotesize{19.72} & \footnotesize{(8)}           & \footnotesize{19.67} & \footnotesize{(12)}  &
\footnotesize{20.06} & \footnotesize{(17)}          & \footnotesize{20.04}       & \footnotesize{(32)}        & \footnotesize{19.71} & \footnotesize{(12)}        \\
\footnotesize{$1\,1\,1$}            &
\footnotesize{22.38} & \footnotesize{(63)}          & \footnotesize{22.53}       & \footnotesize{(25)}        & \footnotesize{22.06} & \footnotesize{(7)}         & \footnotesize{21.98} & \footnotesize{(24)}          & \footnotesize{22.04} & \footnotesize{(6)}   &
\footnotesize{23.89} & \footnotesize{(7)}           & \footnotesize{23.95}       & \footnotesize{(0.1)}       & \footnotesize{22.11} & \footnotesize{(6)}         \\
\footnotesize{$1\,1\,0$}            &
\footnotesize{23.82} & \footnotesize{(19)}          & \footnotesize{23.87}       & \footnotesize{(20)}        & \footnotesize{23.50} & \footnotesize{(56)}        & \footnotesize{23.49} & \footnotesize{(55)}          & \footnotesize{23.47} & \footnotesize{(57)}  &
\footnotesize{24.39} & \footnotesize{(76)}          & \footnotesize{24.39}       & \footnotesize{(83)}        & \footnotesize{23.52} & \footnotesize{(59)}        \\
\footnotesize{$1\,\overline{1}\,0$} &
\footnotesize{32.34} & \footnotesize{(2)}           & \footnotesize{32.31}       & \footnotesize{(7)}         & \footnotesize{31.92} & \footnotesize{(21)}        & \footnotesize{31.97} & \footnotesize{(11)}          & \footnotesize{31.87} & \footnotesize{(31)}  &
\footnotesize{32.16} & \footnotesize{(4)}           & \footnotesize{32.11}       & \footnotesize{(11)}        & \footnotesize{31.92} & \footnotesize{(20)}        \\
\footnotesize{$2\,1\,1$}            &
\footnotesize{35.47} & \footnotesize{(116)}         & \footnotesize{35.64}       & \footnotesize{(116)}       & \footnotesize{34.97} & \footnotesize{(72)}        & \footnotesize{34.89} & \footnotesize{(61)}          & \footnotesize{34.93} & \footnotesize{(69)}  &
\footnotesize{37.20} & \footnotesize{(77)}          & \footnotesize{37.25}       & \footnotesize{(82)}        & \footnotesize{35.03} & \footnotesize{(71)}        \\
\footnotesize{$1\,\overline{1}\,1$} &
\footnotesize{38.30} & \footnotesize{({\bf 100})}   & \footnotesize{38.27}       & \footnotesize{({\bf 100})} & \footnotesize{37.79} & \footnotesize{({\bf 100})} & \footnotesize{37.84} & \footnotesize{({\bf 100})}   & \footnotesize{37.73} & \footnotesize{({\bf 100})} &
\footnotesize{38.19} & \footnotesize{({\bf 100})}   & \footnotesize{38.14}       & \footnotesize{({\bf 100})} & \footnotesize{37.79} & \footnotesize{({\bf 100})} \\
\footnotesize{$2\,1\,0$}            &
\footnotesize{39.68} & \footnotesize{($\approx$ 0)} & \footnotesize{39.75}       & \footnotesize{(0.2)}       & \footnotesize{39.14} & \footnotesize{(2)}         & \footnotesize{39.14} & \footnotesize{(5)}           & \footnotesize{39.08} & \footnotesize{(2)}    &
\footnotesize{40.46} & \footnotesize{(12)}          & \footnotesize{40.45}       & \footnotesize{(8)}         & \footnotesize{39.17} & \footnotesize{(3)}         \\
\footnotesize{$2\,0\,0$}            &
\footnotesize{40.56} & \footnotesize{(4)}           & \footnotesize{40.56}       & \footnotesize{(0.2)}       & \footnotesize{40.03} & \footnotesize{(0.8)}       & \footnotesize{40.05} & \footnotesize{($\approx$ 0)} & \footnotesize{39.95} & \footnotesize{(0.9)} &
\footnotesize{40.77} & \footnotesize{(1)}           & \footnotesize{40.73}       & \footnotesize{(0.8)}       & \footnotesize{40.03} & \footnotesize{(0.9)}       \\
\footnotesize{$2\,2\,1$}            &
\footnotesize{42.34} & \footnotesize{(1)}           & \footnotesize{42.20;42.57} & \footnotesize{(0.2;0.3)}   & \footnotesize{41.73} & \footnotesize{(0.2)}       & \footnotesize{41.62} & \footnotesize{($\approx$ 0)} & \footnotesize{41.68} & \footnotesize{(0.2)}  &
\footnotesize{44.69} & \footnotesize{(2)}           & \footnotesize{44.77}       & \footnotesize{(0.4)}       & \footnotesize{41.82} & \footnotesize{(0.2)}       \\
\hline
\end{tabular}
\parbox{23cm}{
\caption{\label{tab:XRD_theory} XRD theoretical peaks of the (B$_{11}$C$^p$)C$\protect\msquare$C and (B$_{11}$C$^p$)C-C configurations, in the periodically repeated 14-atom ordered elemental unit-cell (columns:~2-3,12-13); in a 378-atom 3x3x3 supercell with substitutional disorder in the polar site of the icosahedra (columns:~4-5,14-15); as an isolated defects in a 404-atom 3x3x3 supercell of B$_4$C$^p$ with polar substitutional disorder (columns:~6-7, 16-17). Peaks of pristine phases are also understood as those of clusters of defects (columns~2-5~and~12-15). Peaks of the B$_4$C$^p$ matrix are given for reference, either for the 15-atom ordered phase (columns:~8-9) or for the phase with polar substitutional disorder (columns:~10-11). 2$\theta$ values are given in Cu~K-$\alpha$. Peak intensities have been normalised to the ($1\,\overline{1}\,1$) peak of each phase. Small peaks ($I$~$<$~4.5) occurring in supercells below 19.56\textdegree~ are not shown.
High 2$\theta$ values have not been observed in present experiments, theoretical ones can be found in table B.13 of Ref.~\protect\cite{Jay:2015} for the ordered phases.
}}
\end{sidewaystable}

\subsection{Formation energy}
\label{subsec:form_ener}
The review of the formation energy of various neutral vacancies in boron carbide shows that the boron atom vacancy at the centre of the C-B-C chains has the lowest formation energy among neutral vacancies~\cite{Vast:2009,Raucoules:2011,Gillet:2018,Roma:2021}. Its value depends on the chemical potential of boron, and ranges from 1.62 to 1.77~eV/defect for the neutral vacancy in a 2x2x2 supercell, and from 2.16 to 2.31~eV/defect for the vacancy charged with one electron (DFT-LDA, Ref.~\cite{Roma:2021}). In our 3x3x3 supercell with substitutional disorder on the polar site, the DFT-LDA formation energy is slightly higher, 1.96~eV for the neutral C$\msquare$C vacancy with respect to B$_4$C with polar disorder (table~\ref{tab:equi_prop_LDA}). For the C-C chain, it amounts to 2.73~eV. We note that the boron chain vacancy is however not the defect having the lowest energy: at low temperature, the Pandey concerted exchange mechanism between the polar carbon atom and the polar boron atom of a neighboring icosahedron leads to the \textit{bipolar} defect complex that has the lowest energy that amounts to only one fourth of an~eV in DFT-LDA~\cite{Mauri:2001,Vast:2009,Roma:2021}.

The fact that vacancies primarily form in the chains is consistent with an earlier study that showed that the boron atom at the chain centre is weakly bonded to the atoms at the chain ends, since their shorter C-B distance, 1.43~{\AA} ~ (see table~\ref{tab:equi_prop_GGA}), points towards substantial $\pi$-bonding~\cite{Balakrishnarajan:2007}. It is also confirmed by more recent calculations of isolated defects and cluster of defects~\cite{Pandey:1986,Raucoules:2011,Betranhandy:2012,Gillet:2018,Roma:2021}.

This is also consistent with experiments: boron atoms at chain centres are known to have a high thermal Debye-Waller factor in X-ray diffraction data. Up to 15-25\% chain centers have been observed to be boron vacancies in neutron diffraction data~\cite{Morosin:1987}. 

\subsection{Atomic structures}
In the absence of the boron atom, the chain evolves from triatomic (C-B-C chains) to the C$\msquare$C configuration, with a drastic change of the first-neighbour distance in the chain $d_{chain}$. Consequent spectroscopic fingerprints of this change can been seen 
in the theoretical XRD peaks (table~\ref{tab:XRD_theory}) and depend on whether the defect is isolated in a matrix of B$_4$C (columns~6-7) with $d_{chain}$~=~3.05~\AA, or whether a large number of neighbouring defects are formed (columns~2-5) with $d_{chain}$~$\approx$~2.86~\AA. 

For an isolated defect, the peak positions are shifted towards slightly higher values than in B$_4$C, with a mean squared shift (MSS) of the nine $2~\theta$ values of 0.017\textdegree. 
The shift is larger for the C-C defect, the MSS reaching 0.027\textdegree. The peak positions are merely shifted towards slightly higher values of $2~\theta$, which reflects a slight crystal volume decrease that is 
larger for the C-C defect than for the C$\msquare$C defect. The diffraction pattern is similar to that of pristine B$_4$C and the occurrence of isolated defects cannot explain the new peaks observed in the experiments. 
The 2~$\theta$ values of isolated defects are very close to those of B$_4$C which indicates that isolated defects, if any, cannot be detected by EDXRD.  

For clustering defects on the contrary, the average theoretical MSS shift of (B$_{11}$C$^{p}$)C$\msquare$C, taken as a crystal, with respect to undeformed (B$_{11}$C$^{p}$)C-B-C, is one order of magnitude larger than that of the isolated defect and amounts to 0.184\textdegree~ (columns~4~and~10). 
The intensities also show some changes (columns~5~and~11), as in particular the relative intensities of the two most intense peaks are modified ($hkl$ peaks~$2\,1\,1$ and~$1\,\overline{1}\,1$).

Changes are more important when chains evolve from triatomic to diatomic (C-C chains) than to C$\msquare$C configurations, with a MSS as large as 0.524\textdegree~ (columns~10~and~14). 

Remarkably, all of the new peaks observed in EDXRD characterisations after large deformation coincide well with the peaks of the (B$_{11}$C$^{p}$)C$\msquare$C, as discussed in section~\ref{sec:disc}. 

\subsection{Behaviour of the chain vacancy under pressure}
\label{subsec:CVC_to_CC}

\begin{figure*}[th!]
\begin{center}
\subfigure[~Defect cluster]  {\includegraphics[clip=true, width=0.49\textwidth, trim=0.3cm 2.0cm 5.1cm 2.5cm]{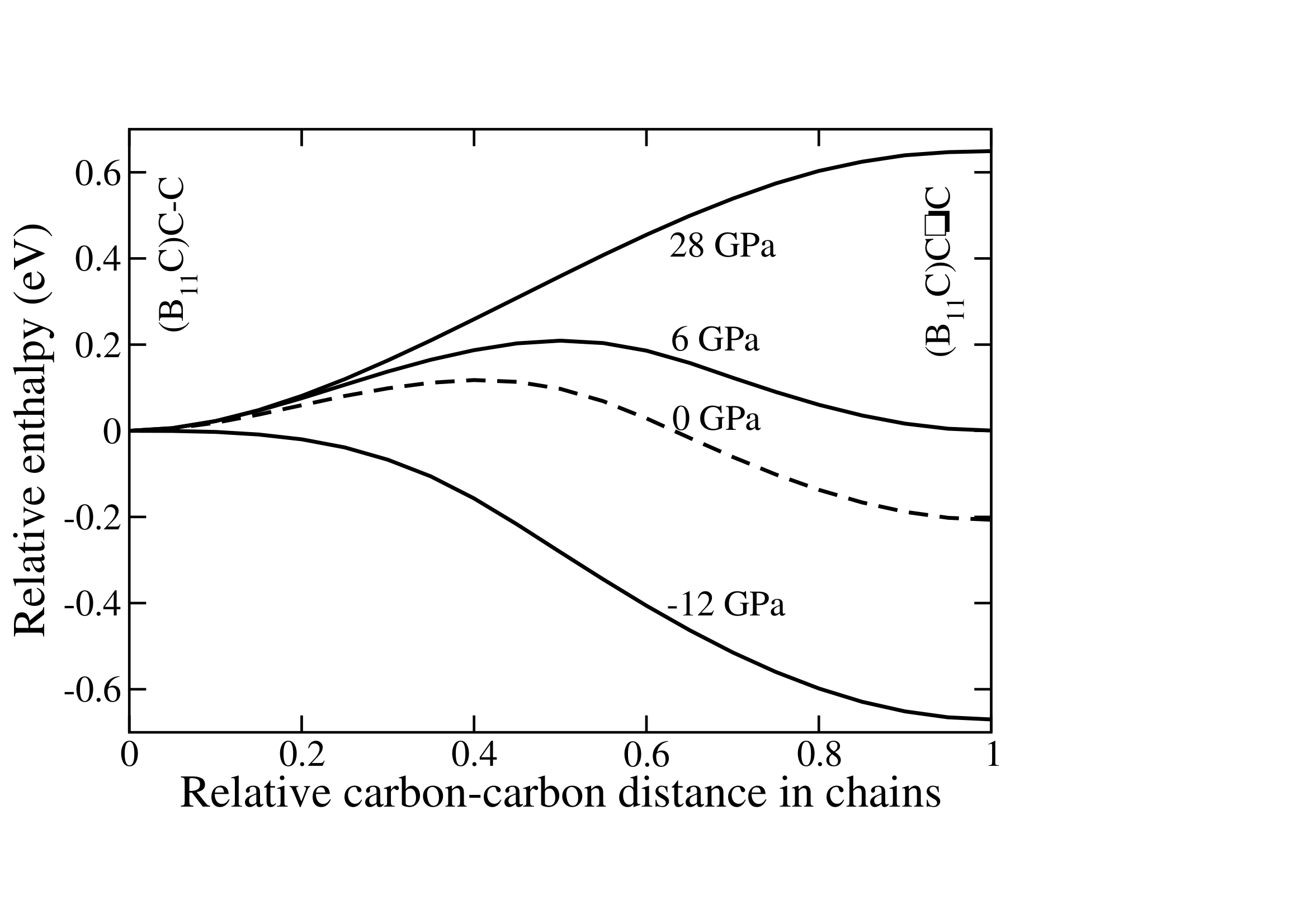} \label{fig:energy_barrier_phase}}
\subfigure[~Isolated defect] {\includegraphics[clip=true, width=0.49\textwidth, trim=0.3cm 2.0cm 5.1cm 2.5cm]{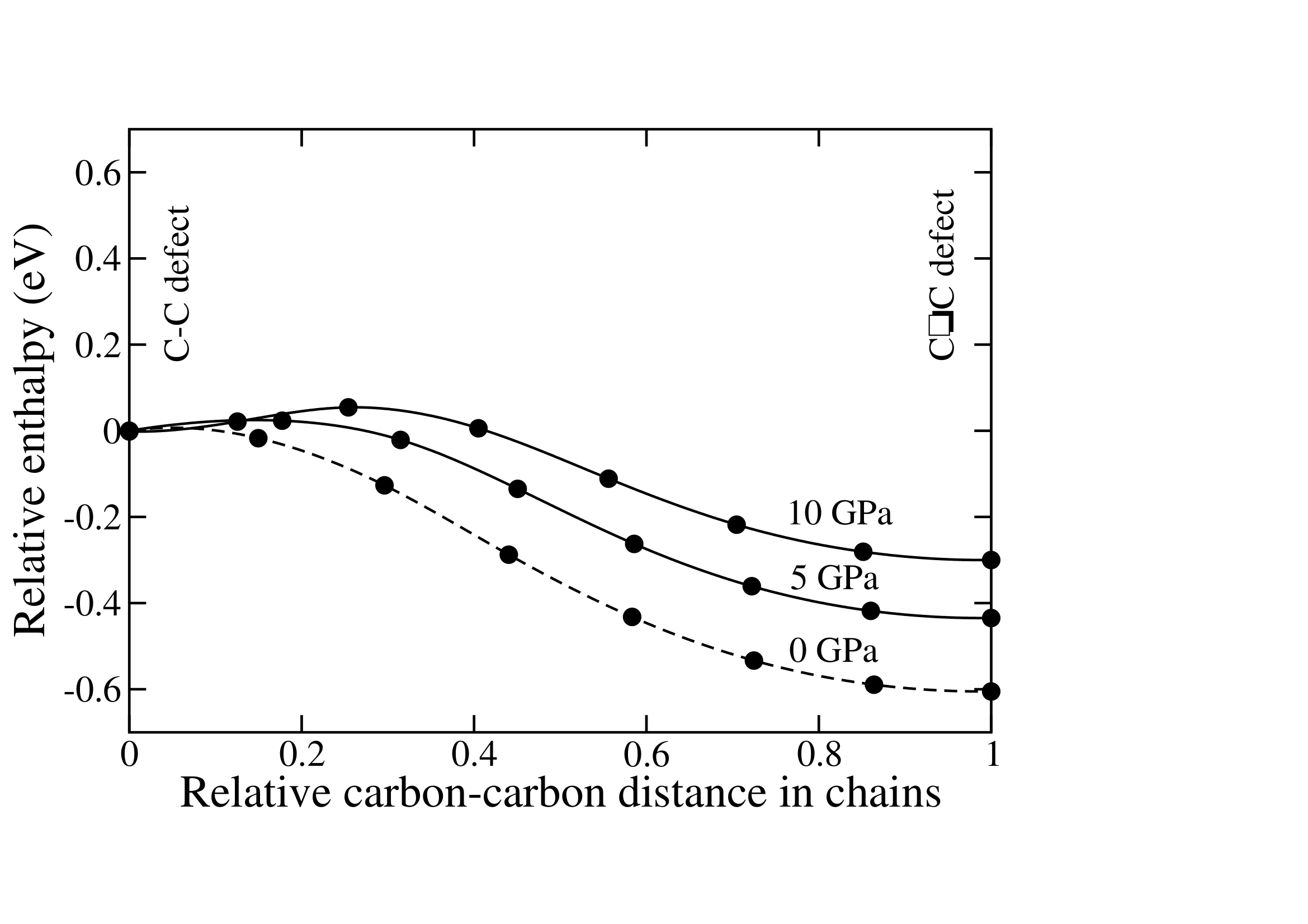}\label{fig:energy_barrier_isolated_defect}}
\caption{\label{fig:energy_barrier} Theoretical energy barrier between the
C$\protect\msquare$C and C-C configurations of the chain. Left panel: in the periodically-repeated 14-atom unit cell, at zero pressure (dashed line), in compression up to 28~GPa, and in extension down to a negative pressure of -12~GPa (solid lines). Volume relaxation was accounted for along the path. Right panel: isolated chain-defect in a 3x3x3 matrix of ordered B$_4$C at ambient pressure (dashed line) and in compression up to 10~GPa (solid lines), computed with the nudge elastic band method at the volume of the B$_4$C matrix for a given pressure. Computations in DFT-GGA-PW91. 
}
\end{center}
\end{figure*}

In this section, we study the effect of pressure on the chain vacancy, as in our experiment, a pressure of 5~GPa is applied simultaneously with the torsional deformation. The question that arose was whether this pressure would be sufficient to modify the vacancy configuration so that a new C-C bond is formed. To this end, we present the enthalpy as a function of the relative distance between the two carbon atoms of the chain for various pressure values (figure~\ref{fig:energy_barrier}).

In fact, the presence of a vacancy in the chain leads to the presence of dangling bonds between the carbon atoms~\cite{Betranhandy:2012}, and to defect-induced energy levels in the band gap of the Kohn-Sham electronic band structure.
Such a defective electronic configuration tends to be restabilised, either, at ambient pressure, by capturing a charge~\cite{Jay:2015,Gillet:2018,Roma:2021}, or, as previously studied~\cite{Raucoules:2011,Betranhandy:2012}, by forming a carbon-carbon bond under pressure (figure~\ref{fig:energy_barrier}).

As far as energy at both ambient pressure and thermodynamical equilibrium is concerned, a crystal in which all of the intericosahedral spaces contain C-C chains
is competing with the crystal in which they all contain C$\msquare$C configurations (figure~\ref{fig:energy_barrier}, left panel, dashed line). The energy difference at ambient pressure between the (B$_{11}$C$^{p}$)C$\msquare$C and (B$_{11}$C$^{p}$)C-C phases is smaller than
100~meV, and it is hard to decide from our calculations which one is the most stable phase: (B$_{11}$C$^{p}$)C$\msquare$C is the most stable phase in DFT-GGA-PW91, while (B$_{11}$C$^{p}$)C-C is the most stable one in LDA (not shown).

At thermodynamical equilibrium the two structures can coexist, and which one of the phase is formed depends on the conditions of formation. The two phases are however separated by an energy barrier of 0.33~eV at zero pressure, so that once vacancies are formed, the probability of formation of C-C bonds is negligible unless one of the physical variables is modified: for instance, application of an hydrostatic pressure lowers the energy barrier. At 6~GPa, the energy barrier is 0.21~eV (figure~\ref{fig:energy_barrier}, left panel). Ultimately, at approximatively 28~GPa in the DFT-GGA-PW91 calculations -and 20~GPa in DFT-LDA~\cite{Betranhandy:2012}- only the (B$_{11}$C$^{p}$)C-C phase can exist. This (hydrostatic) pressure is larger than the expected uniaxial pressure reached in our experiments. 

When the defect is isolated in a matrix of B$_4$C, for steric reasons, the energetic competition turns out to be in favour of the C$\msquare$C configuration (figure~\ref{fig:energy_barrier}, right panel, dashed line). The volume of (B$_{11}$C$^{p}$)C$\msquare$C is smaller then the volume of B$_4$C, the volume of (B$_{11}$C$^{p}$)C-C being even smaller~\cite{Jay:2014} (see table~\ref{tab:equi_prop_GGA} for the volumes of the ordered phases). The presence of the B$_4$C matrix constrains the C$\msquare$C configuration in extension and the C-C chain in even larger extension. The presence of the B$_4$C matrix thus stabilises the C$\msquare$C configuration, by 0.6~eV at ambient pressure (figure~\ref{fig:energy_barrier}, right panel, dashed line). The energy barrier amounts to 0.46~eV at 5~GPa and to 0.35~eV at 10~GPa, so that only under pressures higher than 28~GPa (not shown) does the isolated C-C bond replace the vacancy. 

In conclusion of this section, the formation of a C-C bond, replacing a vacancy, is not expected under a moderate value of the hydrostatic pressure. The case of an uniaxial pressure led to similar conclusions for clusters of defects~\cite{Betranhandy:2012}. Whenever C-C bonds are formed in the experiments, this would be attributed to the unusual conditions in the RotoPEC, where both uniaxial pressure and torsional deformations are applied.

\subsection{Vibrational properties}

The Raman-active vibrational frequencies are reported for the ordered (B$_{11}$C$^{p}$)C-B-C (table~\ref{tab:Ramanpeaks_B4C}), (B$_{11}$C$^{p}$)C$\msquare$C and (B$_{11}$C$^{p}$)C-C phases (table~\ref{tab:Ramanpeaks_new}). 

Going from C-B-C to C-C and to C$\msquare$C configurations in the intericosahedral space induces changes in all of the vibrational frequencies. The main change, however, is for the symmetric stretching mode of the chain ($cs$) that we have called $Ch5$. Its frequency goes from 1095~cm$^{-1}$ to 570~cm$^{-1}$ (in the case of C-C) and 398~cm$^{-1}$ (in the case of C$\msquare$C). This evolution comes from the distance between the chain atoms, which increases from~1.42~\AA ~ to~1.73~\AA ~ and to~2.84~\AA ~ in DFT-LDA (table~\ref{tab:equi_prop_LDA}) (and from 1.43~\AA ~ to 1.80~\AA ~ and to 2.92~\AA ~ in DFT-GGA, see table~\ref{tab:equi_prop_GGA}). 

The libration of the icosahedra has similar frequencies in all of the materials ($Ico5$ also called $lib$ mode) and is present in all of the experiments, which testifies to the presence of icosahedra (resp. 536~cm$^{-1}$, 481~cm$^{-1}$ and 491~cm$^{-1}$ in B$_4$C, (B$_{11}$C$^{p}$)C$\msquare$C and (B$_{11}$C$^{p}$)C-C phases). 

Finally, the (pseudo) chain rotation ($Ico7$ or $pcr$) is one of the characteristic Raman peaks of B$_4$C, together with the librational mode. It is found at 488~cm$^{-1}$ and 529~cm$^{-1}$ in B$_4$C and (B$_{11}$C$^{p}$)C$\msquare$C, and is blue shifted by 140~cm$^{-1}$ between B$_4$C and (B$_{11}$C$^{p}$)C-C, where it is at 621~cm$^{-1}$.

\section{Discussions}
\label{sec:disc}

Both the XRD patterns and Raman spectroscopy of the RP~1 sample show the formation of new peaks. These peaks can be explained by the formation of vacancies in the crystal structure of boron carbide due to the torsional stress generated by the RoToPEC. In order to verify this hypothesis, the changes in the spectra have been compared with the spectra of the various defective boron carbides calculated in section~\ref{sec:theory}. We show that EDXRD and Raman spectroscopy turn out to be complementary and mandatory in characterising the damage formed in the vicinity of the plastic deformation in boron carbide. 

\subsection{Comparison of experimental and theoretical XRD spectra}

All of the four peaks that appear in EDXRD spectra can be explained by the theoretical calculations as coming from the formation of large clusters of unit cells with a boron chain vacancy (table~\ref{tab:EDXRDpeaks}).
The formation of vacancies occurs on distances large enough to be detected by the (very accurate) EDXRD.
The reason behind this cluster formation is attributed to the torsion generated in the RoToPEC.

The two most intense peaks are observed at resp.~37.73\textdegree~ and 34.94\textdegree~ (see also~\ref{app:stress_gradient}). They are attributed to B$_4$C and, remarkably, these two peaks are also predicted to be the most intense by the theoretical calculations for B$_4$C (column~11 of table~\ref{tab:XRD_theory}). They come from the ($1\,\overline{1}\,1$) and ($2\,1\,1$) plans in the trigonal representation of the $R\overline{3}m$ space group.

\begin{table}[th!]
\begin{tabular}{cccccccc}
\hline
Trigonal                & \multicolumn{2} {c}{ EDXRD 2$\theta$(\textdegree)}  & \multicolumn{4} {c}{Theory 2$\theta$(\textdegree) } & Hexagonal \\
\small{$hkl$}           & \small{New}   & \small{Common}& \small{C$\msquare$C}& \small{B$_4$C} & & \small{C-C}   & \small{$hkl$} \\
\hline
\small{$100$}           &               &               & \small{19.87}       & \small{19.67}  & & \small{20.04} & \small{$101$} \\
\small{$111$}           &               & \small{22.04} & \small{22.53}       & \small{22.04}  & & \small{23.95} & \small{$003$} \\
\small{$110$}           &               & \small{23.47} & \small{23.87}       & \small{23.47}  & & \small{24.39} & \small{$012$} \\
\small{$1\overline{1}0$}&               & \small{31.91} & -                   & \small{31.87}  & & \small{32.11} & \small{$110$} \\
\bf{$1\overline{1}0$}   & \small{33.35} &               & \small{32.31}       &  -             & &               & \small{$110$} \\
\small{$211$}           &               & \small{34.94} & -                   & \small{34.93}  & &               & \small{$104$} \\
\bf{$211$}              & \small{36.29} &               & \small{35.64}       & -              & & \small{37.25} & \small{$104$} \\
\small{$1\overline{1}1$}&               & \small{37.73} & \small{38.27}       & \small{37.73}  & & \small{38.14} & \small{$021$} \\
\small{$210$}           &               & \small{39.15} & \small{39.75}       & \small{39.08}  & & \small{40.45} & \small{$113$} \\
\bf{$200$}              & \small{40.75} &               & \small{40.56}       & \small{39.95}  & & \small{40.73} & \small{$202$} \\
\bf{$221$}              & \small{42.51} &               & \small{42.20;42.57} & \small{41.68}  & & \small{44.77} & \small{$015$} \\
\hline
\end{tabular}
\caption{\label{tab:EDXRDpeaks} XRD experimental and theoretical peak positions. 2$\theta$ values are given in Cu~K-$\alpha$.
New peaks that have appeared in the RP~1 experiment at position~15 (figure~\ref{fig:RP1}) and corresponding Miller indices in the trigonal representation of the $R\overline{3}m$~(\#~166) space group; Peaks common to the RP~1 expt. and to undeformed B$_4$C observed at the centre of the sample of the RP~1 expt.; 
Theoretical peaks for (B$_{11}$C$^p$)C$\protect\msquare$C, B$_4$C and (B$_{11}$C$^p$)C-C in the 3x3x3 supercell with substitutional disorder of the carbon atom in the polar site. 
}
\end{table}

The ($2\,1\,1$) peak is computed at 34.93\textdegree~ for B$_4$C and at 35.64\textdegree~ for (B$_{11}$C$^{p}$)C$\msquare$C. These two peaks are theoretically well separated and can indeed be distinguished in the experiments (table~\ref{tab:XRD_theory}). On the contrary, the ($1\,\overline{1}\,1$) peak is the most intense peak for B$_4$C and the second most intense peak for (B$_{11}$C$^{p}$)C$\msquare$C, and are close to each other, at resp.~37.73\textdegree~ and 38.27\textdegree. Only one peak can be seen in the experiment at 37.73\textdegree.

Among the new peaks that appear upon large deformation, two can only be explained by the presence of (B$_{11}$C$^{p}$)C$\msquare$C : they come from the ($1\overline{1}0$) and ($211$)
plans, and are observed at resp. 33.35\textdegree~ and 36.29\textdegree.

The other two new peaks at 40.75\textdegree~ and 42.51\textdegree~ turn out to be in better agreement for (B$_{11}$C$^{p}$)C$\msquare$C than for B$_4$C, 
with computed angles of 40.56\textdegree~ and 42.20\textdegree~ or 42.75\textdegree~ (rather than 39.95\textdegree~ and~41.68\textdegree~ for B$_4$C). 

Finally among the nine lowest ($h\,k\,l$) plans predicted by theory, eight are observed for (B$_{11}$C$^{p}$)C$\msquare$C and B$_4$C, the exception being the lowest ($1\,0\,0$) plan, which is absent from our experiments.
\begin{figure}[th!]
\begin{center}
\includegraphics[width=0.8\textwidth]{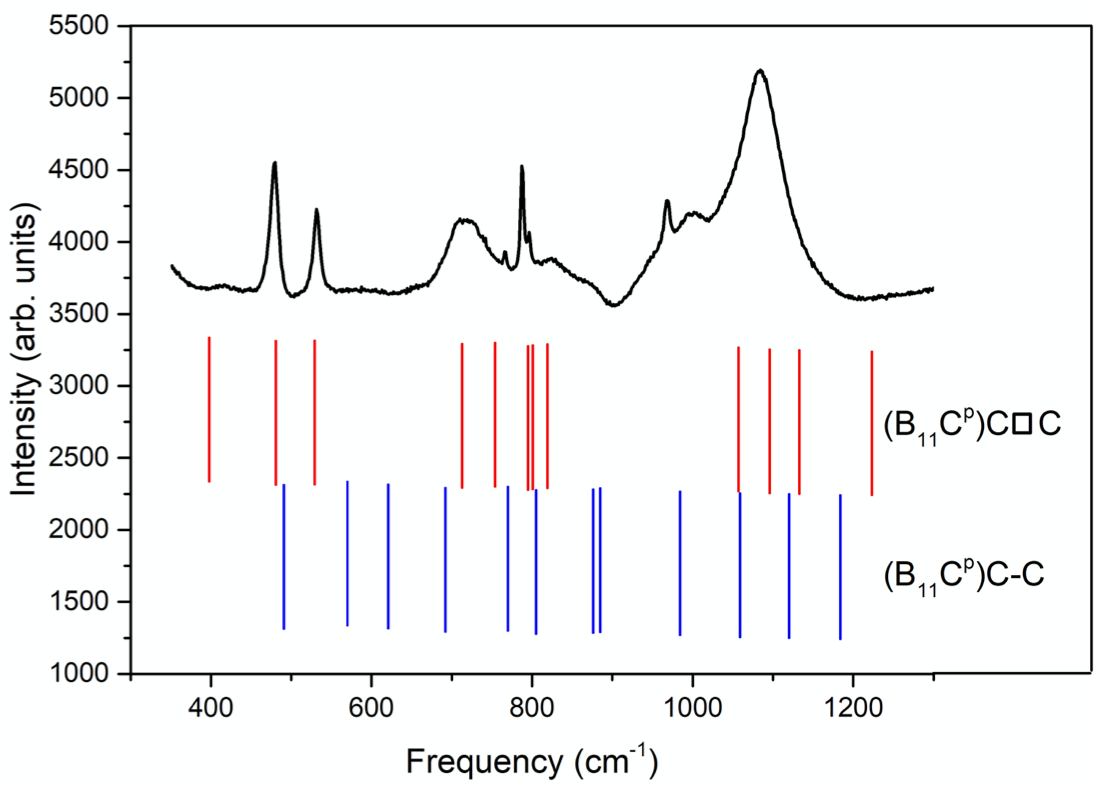}
\end{center}
\caption{\label{fig:Ramanpeaks_new} Experimental Raman spectrum and theoretical Raman-active frequencies. Same spectrum as in figure~\protect\ref{fig:RP1_Raman_1_2} (middle position).
Theoretical frequencies are shown for undeformed (B$_{11}$C$^p$)C$\protect\msquare$C and (B$_{11}$C$^p$)C-C at ambient pressure. 
}
\end{figure}

To quantitatively assess the assignment of the four new peaks, the mean squared error (MSE) of all of the eight~peaks observed at the position~15 of RP~1 that can be corroborated with the peaks of (B$_{11}$C$^{p}$)C$\msquare$C in table~\ref{tab:EDXRDpeaks} has been calculated as
\begin {equation}
MSE=\frac{1}{8} \sqrt { \sum_i (2\theta^{expt} - 2\theta^{theory})^2}
\end{equation}
The MSE amounts to 0.20\textdegree,
for the eight peaks that could correspond to theoretical (undeformed) (B$_{11}$C$^{p}$)C$\msquare$C at ambient pressure (column~4).
Similarly, the mean squared error is 0.15\textdegree~ for the eight peaks that can correspond to the theoretical (undeformed) (B$_{11}$C$^{p}$)C-B-C (theoretical B$_{4}$C) at ambient pressure (column~6).
Both MSEs are on the same order of magnitude, showing that theory yields similar accuracy w.r.t. experiment for (B$_{11}$C$^{p}$)C-B-C and
(B$_{11}$C$^{p}$)C$\msquare$C. We find it to be a strong argument in favour of the presence of (B$_{11}$C$^{p}$)C$\msquare$C in addition to B$_4$C, as the presence of B$_4$C alone is not able to explain all of the observed peaks.

Finally, the EDXRD spectra for RP~2 and RP~3 do not show much change w.r.t. B$_4$C, and this observation underscores the effect of the degree of rotation on the defects formed in the sample - RP~2 (180\textdegree) and RP~3 (90\textdegree) are less deformed compared to RP~1(270\textdegree).
The deformation threshold corresponds to approximately 180\textdegree~ of anvil rotation.

In conclusion, the comparison of EDXRD spectra between theory and experiments leads us to characterise the sample after the largest deformation as multizone, with zones of B$_4$C and zones of (B$_{11}$C$^{p}$)C$\msquare$C, the latter of size smaller than 10~$\mu$m.

\subsection{Comparison of the Raman active frequencies}

The Raman spectrum of distorted B$_4$C is similar to that of the undeformed one (table~\ref{tab:Ramanpeaks_B4C}, columns~2 and~3), and in agreement with the theoretical spectrum (column~4), with the provision that only 8 peaks are singled out in the experiments, as in previous works~\cite{Lazzari:1999,Roma:2021} instead of the expected 12~modes. 
The torsion shifts the peak positions by a few wavenumbers/cm$^{-1}$. 

There are however a number of new peaks which cannot be explained by the presence of B$_4$C only and that we now discuss. They are of two types: peaks attributed to points defects, which turn out to be consistent with the XRD findings of the previous section; and those attributed to amorphous zones that cannot be detected by EDXRD. Thus, EDXRD and Raman spectroscopy complement each other to characterise the damage in B$_4$C.

\begin{table}[th!]
\begin{center}
\begin{tabular}{ccccll}
\hline
\multicolumn{2} {l}{Raman peaks}                 &  \multicolumn{2} {l}{Undeformed B$_{4}$C}     & \multicolumn{2} {l}{Mode description} \\
\footnotesize{New}       & \footnotesize{Common} & \footnotesize{Reference} 
                                                                       & \footnotesize{ Theory}    & \footnotesize{Name}        & \footnotesize{Symmetry}     \\
\hline
\footnotesize{416}&                              &                               &                                                &                                            &                                   \\
                         & \footnotesize{479}    &  \footnotesize{476} & \footnotesize{479-{\bf 488}}  & \footnotesize{$Ico7, pcr$}   & \footnotesize{E$_g$}   \\
                         & \footnotesize{531}    &  \footnotesize{529} & \footnotesize{{\bf 536}-549}  & \footnotesize{$Ico5, lib$}   & \footnotesize{E$_g$}  \\ 
                         &                       &                     & \footnotesize{     699 }      & \footnotesize{$Ico17$}       & \footnotesize{A$_{1g}$} \\ 
                         & \footnotesize{717}    &  \footnotesize{711} & \footnotesize{{\bf 719}}      & \footnotesize{$Ico9$}        & \footnotesize{A$_{1g}$} \\  
\footnotesize{767}       &                       &   -                 & \footnotesize{737-{\bf 768}}  & \footnotesize{$Ico11$}       & \footnotesize{E$_g$}   \\  
\footnotesize{788}       &                       &   -                 &                               &                              &         \\
\footnotesize{798}       &                       &   -                 &                               &                              &         \\
                         &                       &                     & \footnotesize{808-815}        & \footnotesize{$Ico20$}       & \footnotesize{E$_g$}   \\ 
                         &  \footnotesize{823}   & \footnotesize{820}  & \footnotesize{840-841}        & \footnotesize{$Ico13$}       & \footnotesize{E$_g$}   \\ 
                         &  \footnotesize{871}   & \footnotesize{866}  & -                             &                              &        \\
\footnotesize{967$^{*}$} &  \footnotesize{967}   & \footnotesize{964}  & -                             &                              &         \\
                         &  \footnotesize{998}   & \footnotesize{990}  & \footnotesize{994}            & \footnotesize{$Ico21$}       & \footnotesize{A$_{1g}$} \\ 
                         &                       &                     & \footnotesize{{\bf 1052}-1052}& \footnotesize{$Ico23, asis$} & \footnotesize{E$_g$}   \\  
                         &                       &                     & \footnotesize{1069}           & \footnotesize{$Ico24, sis$}  & \footnotesize{A$_{1g}$} \\  
                         & \footnotesize{1082}   & \footnotesize{1079} & \footnotesize{{\bf 1095}}     & \footnotesize{$Ch5, cs$}     & \footnotesize{A$_{1g}$} \\ 
		         &                       &                     & \footnotesize{1098-{\bf 1139}}& \footnotesize{$Ch3, cr$}     & \footnotesize{E$_g$}   \\     
\footnotesize{1198}      &                       &                     &                               &                              &                          \\
\footnotesize{1249}      &                       &                     &                               &                              &                           \\
\footnotesize{1391}      &                       &                     &                               &                              &                              \\

\hline
\end{tabular}
\caption{\label{tab:Ramanpeaks_B4C} Experimental and theoretical Raman-active frequencies (RAF, cm$^{-1}$): 
new peaks that have appeared in the RP~1 experiment, 
RAF common to RP~1 expt. and to distorted boron carbide, RAF of undeformed boron carbide at ambient pressure as observed (column~3) and computed (column~4). The $Cm$ monoclinic symmetry in the calculations slightly lifts the degeneracy of the E$_g$ modes. 
All of the peaks of reference B$_{4}$C are present in the deformed sample. The `*' signifies a much sharper peak at the frequency than the one observed for reference B$_{4}$C. 
$pcr$, $lib$, $asis$, $sis$, $cs$ and $cr$ stand respectively for the pseudo-chain rotation, libration of the icosahedra, anti-symmetric and symmetric intericosahedral stretching, symmetric chain stretching and chain rotation modes. Theoretical frequencies in bold font are expected to be the most intense ones 
}
\end{center}
\end{table}

\subsubsection {Point defects}
In the RP~1 experiment, one new small peak is observed at 416~cm$^{-1}$ or 412~cm$^{-1}$ (figure~\ref{fig:RP1_Raman_2_1}), which corresponds nicely to the $Ch5$ chain stretching ($cs$) mode of the (B$_{11}$C$^p$)C$\protect\msquare$C phase. This is an important finding, as the $Ch5$ mode is expected to be the most intense mode of (B$_{11}$C$^p$)C$\protect\msquare$C (figure~\ref{fig:Ramanpeaks_new}).

Three additional peaks are also observed at resp. 1198, 1249 and 1391~cm$^{-1}$ (figure~\ref{fig:RP1_Raman_1_1}).
The peaks at 1198 and 1249~cm$^{-1}$ correspond well with the peak at 1218~cm$^{-1}$ of the (B$_{11}$C$^p$)C$\protect\msquare$C phase, and with the mode at 1184~cm$^{-1}$ of the (B$_{11}$C$^p$)C-C phase.
However, there has not been any corresponding peaks identified for the last peak at 1391~cm$^{-1}$.

\begin{table}[ht!]
\begin{center}
\begin{tabular}{ccccll}
\hline
\multicolumn{2} {l}{Raman peaks}                 &  \multicolumn{2} {l}{Theory}     & \multicolumn{2} {l}{Mode description} \\
\footnotesize{New}       & \footnotesize{Common} & \footnotesize{(B$_{11}$C$^{p}$)C$\msquare$C}  
                                                                      & \footnotesize{(B$_{11}$C$^{p}$)C-C}  
                                                                                                                        &  \footnotesize{Name}       & \footnotesize{Sym.}     \\ 
\hline
 \footnotesize{416}      &                      & \footnotesize{{\bf 398}}  & \footnotesize{570}            & \footnotesize{$Ch5, cs$}     & \footnotesize{A$_{1g}$} \\              
                         & \footnotesize{479}   & \footnotesize{481-488}    & \footnotesize{410-491}        & \footnotesize{$Ico5, lib$}   & \footnotesize{E$_g$}  \\
                         & \footnotesize{531}   & \footnotesize{529-533}    & \footnotesize{592-621}        & \footnotesize{$Ico7, pcr$}   & \footnotesize{E$_g$}   \\
                         & \footnotesize{717}   & \footnotesize{713}        & \footnotesize{692}            & \footnotesize{$Ico17$}       & \footnotesize{A$_{1g}$} \\ 
                         &                      &                           & \footnotesize{692-708}        & \footnotesize{$Ico13$}       & \footnotesize{E$_g$}   \\ 
                         &                      &                           &                               & \footnotesize{$Ico11$}       & \footnotesize{E$_g$}   \\ 
                         &                      &                           &                               & \footnotesize{$Ico9$}        & \footnotesize{A$_{1g}$} \\ 
\footnotesize{767}       &                      & \footnotesize{730-754}    & \footnotesize{726-770}        & \footnotesize{$Ico11$}       & \footnotesize{E$_g$}   \\ 
\footnotesize{788}       &                      & \footnotesize{795}        & \footnotesize{805}            & \footnotesize{$Ico9$}        & \footnotesize{A$_{1g}$} \\ 
\footnotesize{798}       &                      & \footnotesize{801-827}    &                               & \footnotesize{$Ico20$}       & \footnotesize{E$_g$}   \\   
                         & \footnotesize{823}   & \footnotesize{819-831}    &                               & \footnotesize{$Ico13$}       & \footnotesize{E$_g$}   \\ 
                         & \footnotesize{871}   &                           & \footnotesize{876-918}        & \footnotesize{$Ico20$}       & \footnotesize{E$_g$}   \\     
                         &                      &                           &                               & \footnotesize{$Ico23$}       & \footnotesize{E$_g$}   \\
\footnotesize{967$^{*}$} & \footnotesize{967}   &                           & \footnotesize{984}            & \footnotesize{$Ico21$}       & \footnotesize{A$_{1g}$}   \\
                         & \footnotesize{998}   &                           &                               & \footnotesize{$Ico21$}       & \footnotesize{A$_{1g}$}   \\
                         &                      &                           & \footnotesize{1048-1059}      & \footnotesize{$Ch3, cr$}     & \footnotesize{E$_g$} \\
                         &                      & \footnotesize{1057}       &                               & \footnotesize{$Ico21$}       & \footnotesize{A$_{1g}$} \\
	                 & \footnotesize{1082}  & \footnotesize{1096-1108}  &                               & \footnotesize{$Ico23, asis$} & \footnotesize{E$_g$} \\ 
                         &                      &                           &                               & \footnotesize{$Ico24, sis$}  & \footnotesize{A$_{1g}$}   \\  
\footnotesize{1198}      &                      & \footnotesize{1133}       & \footnotesize{1120}           & \footnotesize{$Ico24, sis$}  & \footnotesize{A$_{1g}$}   \\  
	                 &                      &                           & \footnotesize{885-{\bf 1184}} & \footnotesize{$Ico23, asis$} & \footnotesize{E$_g$} \\ 
\footnotesize{1249}      &                      & \footnotesize{1218-1223}  &                               & \footnotesize{$Ch3, cr$}     & \footnotesize{E$_g$} \\      
\hline
\end{tabular}
\caption{\label{tab:Ramanpeaks_new} Same experimental data as in table~\ref{tab:Ramanpeaks_B4C}, and theoretical Raman-active frequencies (cm$^{-1}$) of undeformed (B$_{11}$C$^p$)C$\protect\msquare$C and (B$_{11}$C$^p$)C-C corresponding to some of the new peaks. The $Cm$ monoclinic symmetry of the former two lifts the degeneracy of the E$_g$ modes. 
Theoretical frequencies in bold font are expected to be the most intense peak for (B$_{11}$C$^p$)C$\protect\msquare$C and (B$_{11}$C$^p$)C-C. 
}
\end{center}
\end{table}

Quite a few new peaks have also appeared at 767, 788, 798, and 967~cm$^{-1}$ in figure~\ref{fig:RP1_Raman_1_2} (middle position). Some of them could be matched to different predicted peaks of (B$_{11}$C$^p$)C$\msquare$C, as shown in figure~\ref{fig:Ramanpeaks_new}. In fact, it is observed from tables~\ref{tab:Ramanpeaks_B4C} and \ref{tab:Ramanpeaks_new} that the entire spectra observed at this middle position could either be explained by peaks of the reference B$_{4}$C or by the peaks of (B$_{11}$C$^p$)C$\msquare$C, which points to the formation of small zones, of size smaller than 2 ~$\mu$m, of (B$_{11}$C$^p$)C$\msquare$C.

However, the peaks corresponding to (B$_{11}$C$^p$)C$\msquare$C that have been noted in figure~\ref{fig:RP1_Raman_1_1} have not appeared in the spectrum of figure~\ref{fig:RP1_Raman_1_2}. Moreover, the peak at 967~cm$^{-1}$ is much sharper than the hump observed in the reference B$_{4}$C around the same frequency. Since this peak does not correspond to any of the (B$_{11}$C$^p$)C$\msquare$C peaks as well, it might be the case that some other polymorph like (B$_{11}$C$^p$)C-C might also be present. Table~\ref{tab:Ramanpeaks_new} shows this possibility of the formation of C-C chains in the sample. This is possible if the C$\msquare$C intericosahedral configurations that had been previously formed under torsion collapsed under the combined torsional and uniaxial deformation to form the C-C chains, as discussed in section~\ref{subsec:CVC_to_CC}. 

The mean square difference between theory and experiment amounts to 5~cm$^{-1}$ for B$_4$C which is remarkable given the fact calculations are limited to the harmonic approximation (table ~\ref{tab:Ramanpeaks_B4C}). It reaches 7~cm$^{-1}$ for (B$_{11}$C$^p$)C$\msquare$C when 11~modes are compared, and
13~cm$^{-1}$ when only the six new modes at resp. 416, 767, 788, 798, 1198, and 1249~cm$^{-1}$ are compared. The MSD for (B$_{11}$C$^p$)C-C amounts respectively to 15~cm$^{-1}$ when 10~modes are compared, and to 19~cm$^{-1}$ when only the six new modes at resp. 416, 767, 798, 967, 1198 and 1249~cm$^{-1}$ are compared.

Finally, in the spectra shown for RP~2 in figure~\ref{fig:RP2_Raman}, there has been no new peaks corresponding to another polymorph of boron carbide, and the spectra obtained for RP~3 in figure~\ref{fig:RP3_Raman} have shown no changes with respect to undeformed boron carbide.

\subsubsection{Amorphous zones}

Another important change is the occurrence, in some of our Raman spectra, of large bands that we tentatively attribute to amorphous zones. 

As explained in the introduction, no clear consensus has been reached so far about the mechanisms leading to amorphisation in boron carbide or about the interpretation of experimental or computational results~\cite{Awasthi:2020}.
Moreover, the kind of amorphous boron carbide obtained in shear bands generated in shock-wave and nanoindentation experiments~\cite{Chen:2003,Yan:2009,Ghosh:2012,Reddy:2013,Subhash:2013} has to be differentiated from the amorphous boron carbide films obtained by chemical vapor deposition~\cite{Pallier:2012,Pallier:2013}. 

In the former, shear bands are associated with two bands centered respectively at 1330 and 1520~cm$^{-1}$, and a small peak at 1810~cm$^{-1}$. 
The first two peaks have been attributed to the $D$ and $G$ bands of amorphous carbon, expected respectively at 1350 and 1580~cm$^{-1}$~\cite{Ghosh:2012,Subhash:2013}, or to a kind of amorphous boron carbide~\cite{Yan:2009,Reddy:2013}. The origin of the peak at 1810~cm$^{-1}$ is not known. DFT calculations cannot explain the occurrence of these peaks unless pressures larger than 60 GPa are achieved~\cite{Awasthi:2020}. 
These shear bands do not occur in nanoindentation experiments when a uniaxial compression smaller than 8~GPa is imposed~\cite{Ghosh:2012}, which is rather far from the value applied in the RoToPEC experiments. 

In the latter, besides the $D$ and $G$ bands of amorphous carbon, two large Raman bands are observed at 400-700~~cm$^{-1}$ and 850-1350~cm$^{-1}$ centered at 1050~cm$^{-1}$~\cite{Pallier:2012,Pallier:2013}. The overall stoichiometry, including free carbon, was B$_{2.5}$C. 

Turning back to our result, a large peak can be seen at 1339~cm$^{-1}$ for instance in figure~\ref{fig:RP1_Raman_2_1}, taken on the sample of the RP~1 experiment. The two other signatures at 1580~cm$^{-1}$ and 1810~cm$^{-1}$ were also present (not shown). We thus tentatively assign these peaks to what is called amorphous boron carbide, as mentioned in the literature about shear bands. 

It is notable that in the present case, the peak that would correspond to amorphous boron carbide has an intensity similar to those of the crystalline boron carbide peaks. In fact, in none of the Raman spectra obtained so far has the amorphous boron carbide peak been more intense than the crystalline boron carbide peaks. Since the laser beam spot of the micro-Raman spectrometer used in these analyses is around 2 $\mu$m, we can safely conclude that the amorphous zones are very localised with areas much smaller than 2 $\mu$m. This is in accordance with what has been reported by Chen~\textit{et al.}~\cite{Chen:2003} in that the bands attributed to amorphous zones, when present, are extremely localised.

Figure~\ref{fig:RP1_Raman_1_1} has been obtained in another direction along the radius of one of the circular faces of RP~~1. A small peak is observed around 1360~cm$^{-1}$ for the spectra taken at the middle of the sample along the radius. This can be attributed to amorphous carbon or graphite (1350~cm$^{-1}$)~\cite{Marton:2013}. This is not unusual as commercial boron carbide is known to contain a certain amount of free carbon in the samples. The latter was however not detected in large quantity in undeformed boron carbide. Further investigation is required to understand whether the $D$ band also comes from the deformation process in the RotoPEC. 

Figure~\ref{fig:RP1_Raman_1_2} has been obtained in yet another direction on the RP~1 sample surface and we observe only a small peak that we tentatively attribute to amorphous boron carbide on the edge, with the caveat given above. 

Finally, in the spectra shown for RP~2 in figure~\ref{fig:RP2_Raman}, one can notice the presence of only a feeble peak attributed to amorphous boron carbide on the edge of the sample.

\subsubsection{Conclusion: Raman spectroscopy}
In conclusion, the comparison of the experimental and theoretical Raman spectra leads us to characterise the sample after the largest deformation as multizone. Apart from B$_4$C in our samples, we also find zones of amorphous boron carbides and/or amorphous carbon, whose size is less than 2~$\mu$m, as well as zones 
of (B$_{11}$C$^{p}$)C$\msquare$C of size less than 2~$\mu$m. We raised the possibility of zones of (B$_{11}$C$^{p}$)C-C. 
The deformation threshold for the occurrence of damage is approximately 180\textdegree~ of anvil rotation, and is detected to be smaller by Raman spectroscopy than by EDXRD.   

\section{Conclusions and perspectives}
\label{sec:conc}

The EDXRD and Raman spectroscopy have clearly provided evidence that chain vacancies 
are produced in boron carbide under the non-hydrostatic torsional stress generated in the RoToPEC, as predicted by Raucoules \textit{et al.}~\cite{Raucoules:2011}. In fact, this is the first experimental proof of chain vacancies forming in boron carbide under non-hydrostatic stress, which can lead to the well-known mechanical failure of boron carbide beyond its Hugoniot elastic limit. 
The samples thus appear as multizone, remaining principally B$_4$C with inclusions of (B$_{11}$C$^{p}$)C$\msquare$C zones and/or amorphous zones. 
The possible presence of other polytypes of boron carbide like (B$_{11}$C$^{p}$)C-C has also been discussed. 
Thus, the present work offers a new solution to the long-standing puzzle of the loss of mechanical strength of boron carbide under dynamic loading.

It also opens the door to several new avenues of further investigation. Both the deformation generated and the defect concentrations can be quantified by repeating the same experiments with a wire inserted at different locations of the sample and noting the deformation of the wire using synchrotron X-ray radiography. Indeed, it is evident from the results that the defects formed depends greatly on the degree of rotation of the gasket, and a quantification of the deformation would also be valuable for further theoretical investigations. 

Obtaining an absolute value of the quantity of defects is a difficult task, and a realistic aim could be to study the defect concentration change as a function of the deformation, with respect to a carefully chosen reference state. The question of the determination of the complicated stress gradient with respect to the radial distance is a possible topic for further investigations in \textit{in situ} experiments in synchrotron facilities. The experiments discussed in this work have been performed at room temperature. That would not result in achieving the HEL in boron carbides, and therefore would not cause any mechanical failure in the samples. However, this will bring boron carbide into the plastic regime, and produce defects and vacancies similar to those produced in boron carbides as it approaches the HEL, thus giving an opportunity to study these defects in a controlled and systematic manner. Higher temperatures can be used in later experiments to approach the HEL more closely and, eventually, investigate the probable effect of defect annealing.

Finally, we point out that EDXRD and Raman spectroscopy are complementary, and in our view both of them should systematically be used in the study of damage in boron carbide. 
The probable occurrence of amorphous zones could be detected, however further theoretical and experimental investigations are required to understand their formation mechanism(s).

\section*{Acknowledgement}
Supports from the DGA (France) and from the program NEEDS-Matériaux (France) are gratefully acknowledged. The authors thank Benoit Baptiste, 
Ludovic Delbes, 
Hicham Moutaabbid, and Silvia Pandolfi for useful discussions. The authors also thank Dr. Nicolas Guignot of PSICHE beamline in SOLEIL synchrotron. The Raman spectroscopy platform in IMPMC is also acknowledged. The PhD fellowship for A. Chakraborti has been provided by the Ecole Doctorale of Institut Polytechnique de Paris.




 \bibliographystyle{elsarticle-num} 
 \bibliography{ref.bib}

\begin{thebibliography}{10}
\bibitem{Chakraborti:Note:2021:defect_annealing}
We note that a possible annealing of defects due to the elevated temperature in
  the shock wave experiments has not been considered in the above estimation of
  the pressure.
\end{thebibliography}

\appendix

\section{Equilibrium properties of the three structural models}
\label{app:structural_model_details}

Equilibrium properties of the three kinds of model used in the present work are reported for DFT-GGA-PW91 (table~\ref{tab:equi_prop_GGA}) and DFT-LDA (table~\ref{tab:equi_prop_LDA}).

The formation energy of the ordered (O) 14-atom (B$_{11}$C$^p$)C$\msquare$C and(B$_{11}$C$^p$)C-C phases is slightly smaller than that of the corresponding models with substitutional disorder (D) of the carbon atom in the six equivalent positions of the polar site of the icosahedra. We note that the substitutional disorder is responsible for the fact that the unit-cell is on average trigonal with the $R\overline{3}m$ symmetry. The formation energy of one isolated defect (ID) in a matrix of B$_4$C with polar substitutional disorder is also given in the tables, and discussed in the main text (section~\ref{subsec:form_ener}).

The equilibrium geometry shown in tables~\ref{tab:equi_prop_GGA}~and~\ref{tab:equi_prop_LDA} consists of the lattice parameters of the base-center monoclinic cell, defined by the unit cell lengths $a_{bcm}$, $b_{bcm}$ and angles $\alpha_{bcm}$ and $\beta_{bcm}$. Their mean trigonal value has been computed as described below, in~\ref{app:average_trigonal_unit_cell}, and the trigonal length $a_t$ and trigonal angle $\alpha_t$ are given.
The equilibrium geometry is also reported for the 3x3x3 supercells with polar disorder. 
The equilibrium volume and atomic positions of supercells with one isolated defect have been relaxed, contrarily to those of figure~\ref{fig:energy_barrier_isolated_defect}, where they have been fixed to those of the B$_4$C matrix.

Finally, the distance between the carbon atom at one of the chain ends and its nearest neighbour in the chain (C-B or C-C distance) is reported after relaxation of the unit-cell and of the atomic positions.

\begin{table}[th!]
\begin{tabular}{lcccccccc}
\hline
                    &\multicolumn{3} {c}{\small{C$\msquare$C}}  & \multicolumn{2} {c}{\small{B$_4$C} matrix}   &  \multicolumn{3} {c}{\small{C-C}}        \\
Equi. Prop.         & \small{O} & \small{D} & \small{ID}
                    & \small{O} & \small{D} & 
                      \small{O} & \small{D} & \small{ID}  \\
\hline
$E_f$               \small{(meV/at.)}      &  -55   & -56     &         & -148  & -139      &   -40  & -38    &        \\ 
$E^d_f$         \small{(eV/defect)}        &        &         & 1.79    &       &           &        &        & 2.73   \\
$V$                  \small{(\AA$^3$)}     & 105.1  &  104.7  & 109.8   & 109.4 & 109.9     & 100.1  & 99.8   & 109.6   \\
$a_{bcm}$            \small{(\AA)}         & 5.135  &         &         & 5.219 &           & 4.920  &        &        \\
$b_{bcm}$            \small{(\AA)}         & 5.011  &         &         & 5.076 &           & 4.902  &        &        \\
$a_t$                \small{(\AA)}         & 5.093  & 5.080   & 5.170   & 5.171 & 5.171     & 4.920  & 4.91   & 5.163  \\
$\alpha_{bcm}$       \small{(\textdegree)} & 66.20  &         &         & 65.98 &           & 68.93  &        &        \\      
$\beta_{bcm}$        \small{(\textdegree)} & 64.87  &         &         & 64.45 &           & 68.86  &        &        \\
$\alpha_t$           \small{(\textdegree)} & 65.76  & 66.10   & 65.72   & 65.47 & 65.72     & 68.91  & 69.08  & 65.81 \\ 
$d_{chain}$          \small{(\AA)}         &  2.92  & 2.86    & 3.05    & 1.43  & 1.43      & 1.80   & 1.77   & 1.71   \\
\hline
\end{tabular}
\caption{\label{tab:equi_prop_GGA}
Formation energy $E_f$, defect formation energy $E^d_f$, unit-cell volume $V$, base-centered monoclinic lattice parameters $a_{bcm}$, $b_{bcm}$, $\alpha_{bcm}$ and $\beta_{bcm}$, average trigonal lattice parameters $a_t$ and $\alpha_t$, and distance $d_{chain}$ between the chain-end carbon atom and its nearest neighbour in the chain of the (B$_{11}$C$^p$)C$\protect\msquare$C and (B$_{11}$C$^p$)C-C configurations, in the three models used in the present work. In a periodically repeated 14-atom ordered (O) elemental unit-cell (columns:~2~and~7); in a 378-atom 3x3x3 supercell with substitutional disorder (D) in the polar site of the icosahedra (columns:~3~and~8); for an isolated defects (ID) in a 404-atom 3x3x3 supercell of B$_4$C$^p$ with polar substitutional disorder (columns:~4~and~9). Supercell volumes and cell lengths have been normalized to the elementary unit- cell to enable comparison. 
Equilibrium properties of pristine phases are also understood as those of clusters of defects (columns~2-3~and~7-8). Peaks of the B$_4$C$^p$ matrix are given for reference, either for the 15-atom ordered phase (column~5) or for the phase with polar substitutional disorder (column~6). Calculations in DFT-GGA-PW91. 
}

\end{table}

\begin{table}[th!]
\begin{tabular}{lccccccc}
\hline
                    & \multicolumn{2} {c}{\small{C$\msquare$C}}  & \multicolumn{2} {c}{\small{B$_4$C} matrix}   &  \multicolumn{3} {c}{\small{C-C}}        \\
Equi. Prop.                & \small{O} &  \small{ID}
                    & \small{O} & \small{D} & 
                      \small{O} & \small{D} & \small{ID}  \\
\hline
$E_f$  \small{(meV/at.)}              &  -14   &         & -121  &  -111     &   -25  & -21    &        \\
$E^d_f$ \small{(eV)}                  &        & 1.96    &       &           &        &        & 2.57   \\
$V$ \small{(\AA$^3$)}                 & 100.6  & 105.2   & 104.8 & 105.4     &  95.60 & 95.38  & 105.0  \\
$a_{bcm}$ \small{(\AA)}               & 5.055  &         & 5.145 &           & 4.845  &        &        \\
$b_{bcm}$ \small{(\AA)}               & 4.930  &         & 4.997 &           & 4.827  &        &        \\
$a_t$  \small{(\AA)}                  & 5.013  & 5.093   & 5.096 & 5.10      & 4.845  & 4.84   & 5.087  \\
$\alpha_{bcm}$ \small{(\textdegree)}  & 66.37  &         & 66.09 &           & 69.00  &        &        \\
$\beta_{bcm}$  \small{(\textdegree)}  & 65.05  &         & 64.41 &           & 68.97  &        &        \\
$\alpha_t$ \small{(\textdegree)}      & 65.95  &  65.80  & 65.53 & 65.79     & 69.00  & 69.16  & 65.88 \\
$d_{chain}$ \small{(\AA)}             &  2.84  & 2.97    & 1.42  & 1.42      & 1.73   & 1.71   & 1.65   \\
\hline
\end{tabular}
\caption{\label{tab:equi_prop_LDA} 
Same as table~~\ref{tab:equi_prop_GGA} for calculations performed in DFT-LDA used for Raman shifts, with the exception of disordered (B$_{11}$C$^p$)C$\protect\msquare$C. 
}
\end{table}

\section{Average trigonal unit cell for ordered boron carbide}
\label{app:average_trigonal_unit_cell}

In the $C_m$ space group of ordered boron carbide, only two symmetry operations leave the crystal invariant:
\begin{itemize}
\item 1 : Identity.
\item 2 : The reflection w.r.t. the (101) mirror plan passing through the chain and the carbon atom in the polar site of the icosahedron.
\end{itemize}

In order to obtain average trigonal parameters from the monoclinic cell parameters of the ordered unit cells, $a_t$ et $\alpha_t$,
the following relationships have been used (see Ref.~\cite{Jay:2015} for details): 
\begin{equation}
a_t=\frac{a_{bcm}+b_{bcm}+c_{bcm}}{3}, 
\end{equation}
\begin{equation}
\alpha_t=\frac{\alpha_{bcm}+\beta_{bcm}+\gamma_{bcm}}{3},
\end{equation}
where $c_{bcm}$=$a_{bcm}$ and $\gamma_{bcm}=\alpha_{bcm}$. 

\begin{table}[!ht]
\begin{center}
\begin{tabular}{lcccccc}
\hline
\hline
Site common name     & Multiplicity& L. & Sym.& \multicolumn{2}{c}{Atomic position} \\
\hline
Chain center         & 1     & a  & $m$ & $x,y,x$           \\
Chain end 1          & 1     & a  & $m$ & $x,y,x$           \\
Chain end 2          & 1     & a  & $m$ & $x,y,x$           \\
Polar site 1         & 1     & a  & $m$ & $x,y,x$           \\
Polar site 2         & 1     & a  & $m$ & $x,y,x$           \\
Polar site 3         & 2     & b  & $1$ & $x,y,z$ & $z,y,x$ \\
Polar site 4         & 2     & b  & $1$ & $x,y,z$ & $z,y,x$ \\
Equatorial site 1    & 1     & a  & $m$ & $x,y,x$           \\
Equatorial site 2    & 1     & a  & $m$ & $x,y,x$           \\
Equatorial site 3    & 2     & b  & $1$ & $x,y,z$ & $z,y,x$ \\
Equatorial site 4    & 2     & b  & $1$ & $x,y,z$ & $z,y,x$ \\
\hline
\hline
 \end{tabular}
\end{center}
\caption{\label{Wickoff_pos_mono} The Wyckoff sites of theoretical base-centered ordered boron carbide: multiplicity, Wyckoff's label (L.), site symmetry, and atomic coordinates according to the international table of crystallography~\cite{Tablecristallo}. 
}
\end{table}

To obtain atomic positions in the average trigonal unit cell, at the Wyckoff sites $1b$, $2c$, $6h$ et $6h$ of the $R\overline{3}m$ space group, 
atomic positions that have the same common name in table~\ref{Wickoff_pos_mono} have been averaged with the following relationships: 
\begin{center}
\begin{eqnarray}
\text{Chain center $1b$:} \nonumber & \\
x=y=z=\frac{1}{2}  &\\
\text{Chain end $2c$:} \nonumber & \\
x=y=z=\frac{1}{6}\sum_{i=2}^3{2x_i+y_i}  & \\
\text{Polar site $6h$:} \nonumber & \\
x=z=\frac{1}{12}\left(\sum_{i=6}^7{x_i+z_i} +2\sum_{i=10}^{11}{x_i+z_i}\right) \nonumber & \\
y=\frac{1}{6}\left(\sum_{i=6}^7{y_i} +2\sum_{i=10}^i{11}{y_i}\right) & \\
\text{Equatorial site $6h$:} \nonumber & \\
x=z=\frac{1}{12}\left(\sum_{i=4}^5{x_i+z_i} +2\sum_{i=8}^9{x_i+z_i}\right)  \nonumber & \\
y=\frac{1}{6}\left(\sum_{i=4}^5{y_i} +2\sum_{i=8}^9{y_i}\right) 
\end{eqnarray}
\end{center}
where $i$ is the atom index and $x$,$y$ and $z$ are atomic coordinates in the crystal framework, along the Bravais lattice vectors. 

In the formula for sites $6h$, one has to account for circular permutations and sign changes of coordinates as given in Ref.~\cite{Tablecristallo}: 
$x,x,z$; $x,z,x$; $z,x,x$; $\overline{x},\overline{x},\overline{z}$; $\overline{x},\overline{z},\overline{x}$; $\overline{z},\overline{x},\overline{x}$. 

The XRD patterns have been computed with these symmetrized lattice parameters and atomic positions for ordered boron carbide. 

\section{Modelling of substitutional disorder}
\label{app:modelling_substitutional_disorder}

\begin{figure}[th!]
\begin{center}
\includegraphics[width=0.5\textwidth]{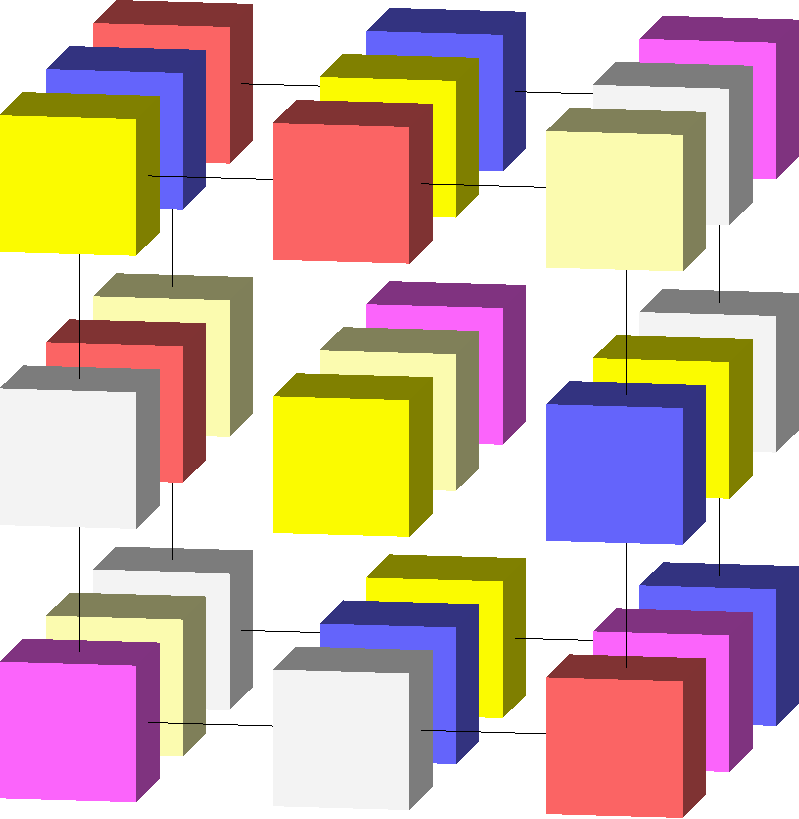}
\end{center}
\caption{\label{fig:substitutional_disorder} Schema of the 3x3x3 supercell used for the modelling of the substitutional disorder of the polar carbon atom in the six equivalent atomic positions of the 6$h$ Wyckoff site of the trigonal $R\overline{3}m$ elemental unit cell. One colour corresponds to one of the six equivalent atomic positions.}
\end{figure}

To model the substitutional disorder of the carbon atom in the icosahedra, a 3x3x3 supercell containing 27 icosahedra and chains has been used. This supercell allows the (quasi) equiprobable distribution of the carbon atom in each of the six equivalent atomic positions of the polar site. Intericosahedral C-C bonds are avoided as the total energy turns out to be increased for such configurations~\protect\cite{Jay:2019}.

\section{Experimental determination of the stress gradient}
\label{app:stress_gradient}

In figure~\ref{fig:RP1_Raman_1_2}, the new peaks have appeared on the middle of the radial distance and not, as was expected, on the edge. This leads to the hypothesis that the defects formed were not a direct function of radial distance, contrarily to what was expected. In order to verify this, the two most prominent peaks of the boron carbide EDXRD spectra, peaks corresponding to (021) and (104) plans in the hexagonal representation of the trigonal structure, have been followed exhaustively to note the change in peak position and the FWHM (Full width at half maximum) with respect to the reference B$_{4}$C powder at ambient pressure. As the relative peak position and FWHM is proportional to the deformation produced, it would allow one to determine if the defects are produced as a direct function of radial distance. Figures are shown here as examples to prove that this is not the case - the relative peak positions and FWHM have no simple direct relationship with the radius. It is possible that the defects have been generated as a function of radial distance and then, there were relaxation of the defects over time, which led to the complicated strain gradient that we see now. It would be useful to follow the evolution of the defects generated over time through \textit{in situ} RoToPEC experiments in the SOLEIL synchrotron, to see if that is indeed the case. It is however difficult to verify that in post mortem characterisations like has been done in the present study.

\begin{figure}[th!]
\begin{center}
\includegraphics[width=0.6\textwidth]{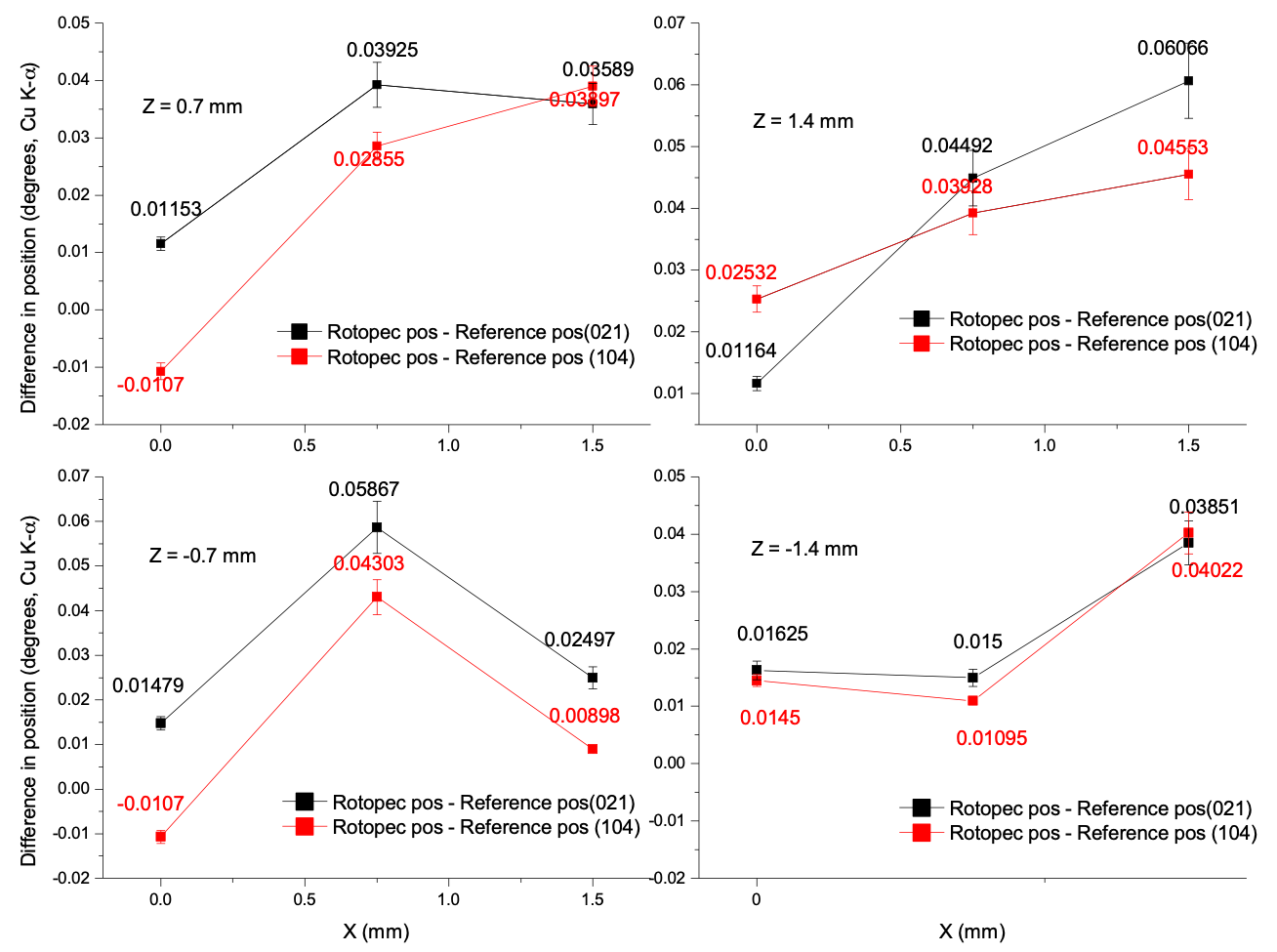}
\end{center}
\caption{\label{fig:peakposition} Analysis of the shift in XRD peak positions of the (021) and (104) plans in the hexagonal representation of the trigonal structure of the deformed boron carbide sample RP~1 (rotated by~270\textdegree) with respect to the reference boron carbide powder at ambient pressure. $z$ represents the height of the particle position from the centre of the rectangular cross-section, as shown in figure~\protect\ref{XRDpositions}.}
\end{figure}

\begin{figure}[H]
\begin{center}
\includegraphics[width=0.6\textwidth]{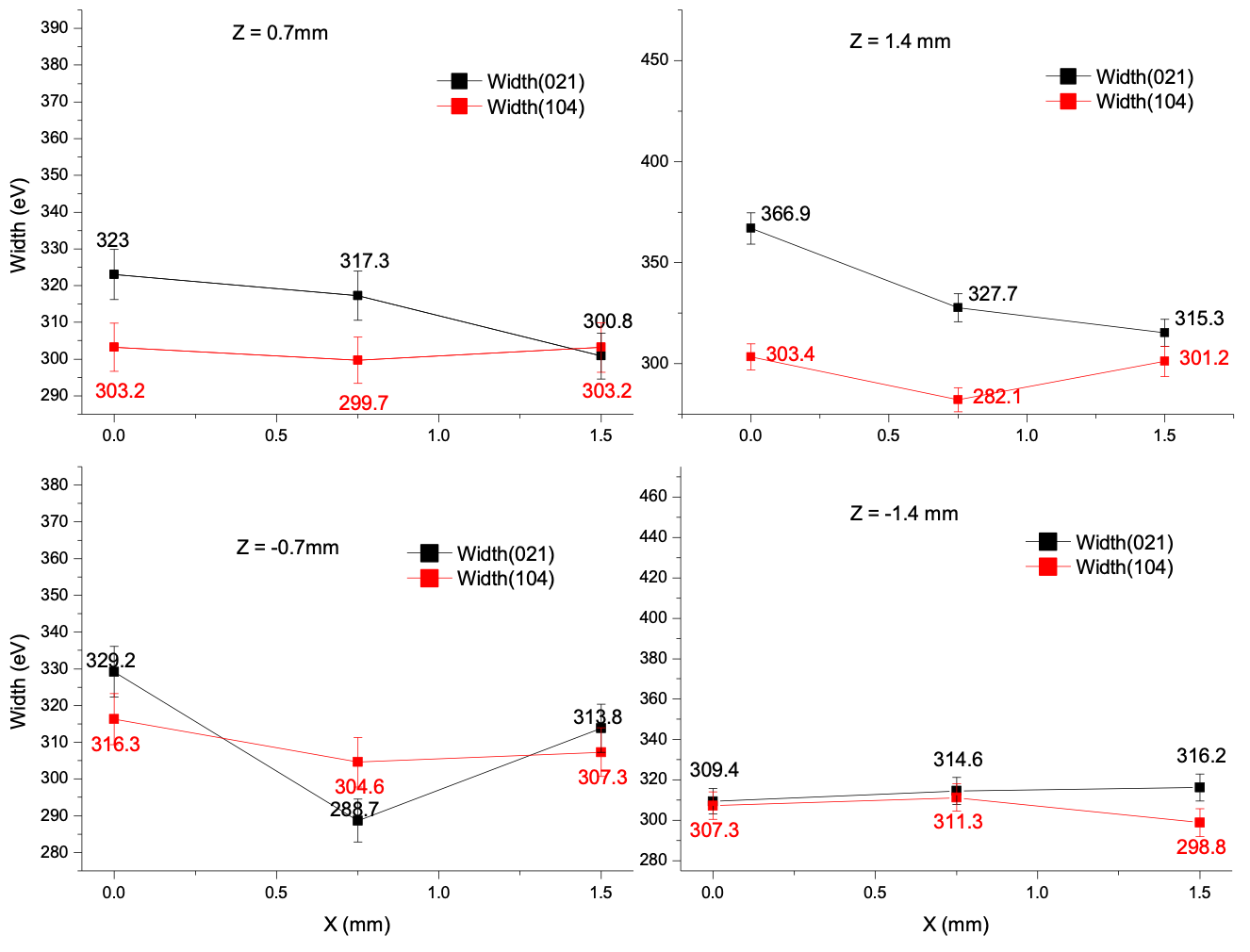}
\end{center}
\caption{\label{fig:width}Analysis of the shift in the FWHM of the XRD peaks of the (021) and (104) plans in the hexagonal representation of the trigonal structure of the deformed boron carbide sample RP~1 (rotated by~270\textdegree) at ambient pressure with respect to the reference boron carbide powder. $z$ represents the height of the particle position from the centre of the rectangular cross-section, as shown in figure~\protect\ref{XRDpositions}. }

\end{figure}




\end{document}